\documentclass[5p]{elsarticle}

\pdfoutput=1
\usepackage{amsmath,amsfonts,amssymb,bm,graphicx,color}

\usepackage{hyperref}
\usepackage{color}

\newcommand{\changes}[1]{{\color{black}{#1}}}

\renewcommand{\L}{\mathcal{L}}
\newcommand{\W}{\mathcal{W}}

\newcommand{\I}{\mathbb{I}}

\newcommand{\er}[1]{Eq.\,\eqref{#1}}
\newcommand{\ers}[2]{Eqs.\,(\ref{#1}-\ref{#2})}
\newcommand{\era}[2]{Eqs.\,(\ref{#1}) and (\ref{#2})}

\newcommand{\Er}[1]{Equation \eqref{#1}}
\newcommand{\Ers}[2]{Equations (\ref{#1}-\ref{#2})}

\newcommand{\ket}[1]{|#1\rangle}

\newcommand{\Tr}{\mathop{\mathrm{Tr}}}

\newcommand{\fs}{\langle - |}

\newcommand{\norm}[1]{\lVert#1\rVert}

\begin{document}

\title{Aspects of non-equilibrium in classical and quantum systems: \\
Slow relaxation and glasses, dynamical large deviations, quantum non-ergodicity, and open quantum dynamics}

\author{Juan P. Garrahan}
\address{School of Physics and Astronomy, \\ 
and 
Centre for the Mathematics and Theoretical Physics of Quantum Non-equilibrium Systems, \\ University of Nottingham, Nottingham NG7 2RD, United Kingdom}

\begin{abstract}
In these four lectures I describe basic ideas and methods applicable to both classical and quantum systems displaying slow relaxation and non-equilibrium dynamics. The first half of these notes considers classical systems, and the second half, quantum systems. In Lecture 1, I briefly review the glass transition problem as a paradigm of slow relaxation and dynamical arrest in classical many-body systems. I discuss theoretical perspectives on how to think about glasses, and in particular how to model them in terms of kinetically constrained dynamics. In Lecture 2, I describe how via large deviation methods it is possible to define a statistical mechanics of trajectories which reveals the dynamical phase structure of systems with complex relaxation such as glasses.  Lecture 3 is about closed (i.e. isolated) many-body quantum systems. I review thermalisation and many-body localisation, and consider the possibility of slow thermalisation and quantum non-ergodicity in the absence of disorder, thus connecting with some of the ideas of the first lecture. Lecture 4 is about open quantum systems, that is, quantum systems interacting with an environment. I review the description of open quantum dynamics within the Markovian approximation in terms of quantum master equations and stochastic quantum trajectories, and explain how to extend the dynamical large deviation method to study the statistical properties of ensembles of quantum jump trajectories.  My overall aim is to draw analogies between classical and quantum non-equilibrium and find connections in the way we think about problems in these areas.
\end{abstract}

\maketitle

\tableofcontents

\section*{Introduction}

The purpose of these lecture notes is to introduce some general and simple ideas about slow relaxation and non-equilibrium dynamics in many-body systems, both classical and quantum.  The aim is not to be comprehensive, but rather to describe particular ways in which to address certain interesting questions in non-equilibrium, and to highlight potential connections between problems in areas that may appear very different.  The first part of these notes deals with classical systems. Lecture \ref{Lec1} is about the glass transition problem, an important and yet not fully understood general problem in condensed matter science, and also a paradigm of slow and complex relaxation more generally. I describe basic questions that emerge from the phenomenology of glass forming systems such as supercooled liquids, and briefly discuss basic theoretical perspectives. Most of the focus is on a general modelling of glasses in terms of systems with constraints in their dynamics, an approach that has wider applicability, as discussed later in the notes.  A key insight that emerges from these considerations is that the interesting behaviour in many  systems with cooperative dynamics is to be encountered in properties of the trajectories of the dynamics rather than in configurations, highlighting the need for a statistical mechanics approach to study trajectory ensembles. Lecture \ref{Lec2} describes how such approach can be constructed with dynamical large deviation methods. Such an approach leads to thinking about dynamics in a thermodynamic-like fashion, for example by revealing the existence of dynamical phases - and phase transitions between them - that underly observed fluctuation behaviour in the evolution of systems with cooperative and collective dynamics. 

The second part of these notes is about non-equilibrium in quantum systems.  Lecture \ref{Lec3} discusses dynamics in isolated quantum many-body systems, where despite unitary evolution, there is both equilibration and thermalisation.  I furthermore describe many-body localisation in disordered systems as a novel paradigm for quantum non-ergodicity. I also consider similarities and differences between many-body localisation and slowdown and arrest in classical glasses, contrasting mechanisms based on disorder to those based on dynamical constraints.  Lecture \ref{Lec4} is about open quantum systems.  I discuss how the dynamics of quantum systems that interact with an environment can be described in an approximate Markovian way, considering similarities and differences with classical stochastic systems.  I also explain how to extend large deviation methods to the open quantum case in order to study properties of quantum jump trajectories using similar ideas to those employed in classical non-equilibrium.

There are many excellent reviews on several of the topics covered in these notes.  For the glass transition problem these include Refs.\ \cite{Binder2011,Ediger1996,Cavagna2009,Berthier2011,Biroli2013,Lubchenko2007,Chandler2010}; for kinetically constrained models, Refs.\ \cite{Ritort2003,Garrahan2010}; for large deviations, Refs.\ \cite{Touchette2009,Touchette2011,Touchette2017}.  In the case of quantum systems, comprehensive recent reviews on thermalisation and many-body localisation include Refs.\ \cite{DAlessio2016,Gogolin2016,Nandkishore2015}; and on open quantum systems, Refs.\ \cite{Plenio1998,Breuer2002,Gardiner2004b,Daley2014}.
The selection of topics, many of the examples shown, and the overall approach to  the problems discussed here is also based on my own work in these areas.

\begin{figure*}[t]
\begin{center}
	\includegraphics[width=1.9\columnwidth]{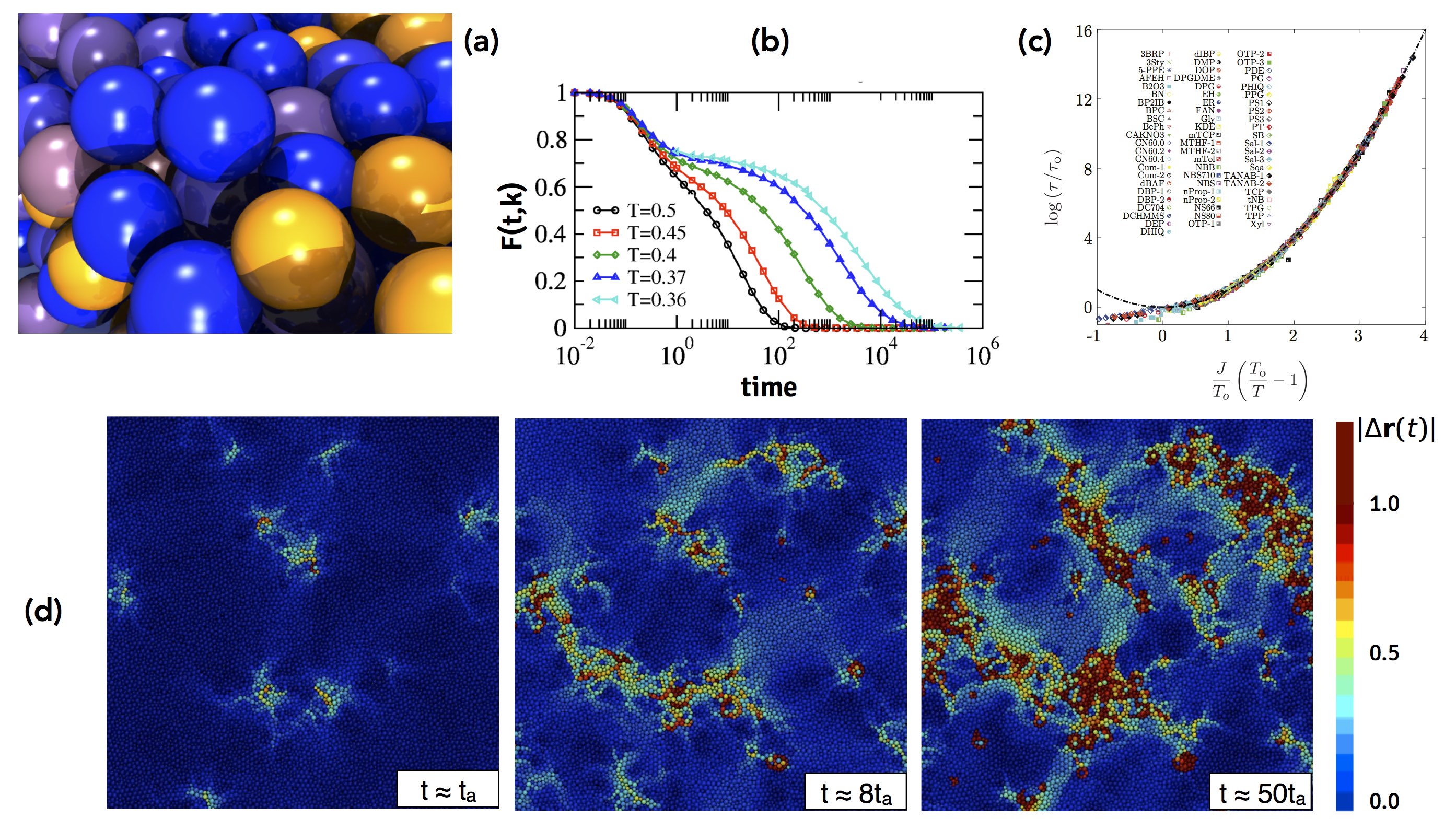}
\caption{{\bf Glass transition:}
(a) Sketch of a system of hard spheres with excluded volume interaction at high density. 
(b) Self-intermediate scattering function of a Lennard-Jones like 3D mixture as a function of time, at various temperatures $T$ within the supercooled regime, for $k$ corresponding to the peak of the static structure factor, from molecular dynamics simulations.  At lower temperatures there is separation of time scales and metastability. From Ref.\ \cite{Chandler2006}.
(c) Relaxation time (or viscosity) as a function of inverse temperature for a number of supercooled liquids from experimental data. Relaxation times are super-Arrhenius.  By convention, when $\tau = 100s$ the experimental glass transition ensues.  Plot shows that for all these liquids the low temperature behaviour is well described by a ``parabolic law'', $\ln \tau = \left[ {J} \left( {T_{\rm o}}/{T} - 1 \right) / {T_{\rm o}} \right]^2 + \ln \tau_{\rm o}$, which in contrast to the traditional VFT form does not predict a finite temperature singularity. From Ref.\ \cite{Elmatad2009}.
(d) Pattern of spatial relaxation showing dynamic heterogeneity. From Ref.\ \cite{Keys2011}.
}
\label{fig1}
\end{center}
\end{figure*}

\section{Slow relaxation in classical systems}
\label{Lec1}

\subsection{Phenomenology of the glass transition}

In the physical sciences glasses are the paradigm of non-equilibrium matter: when too cold or too dense fluids cease to flow, forming the amorphous solid-like material we call glass. This solidification occurs in the absence of any apparent structural ordering, in contrast to more conventional condensed matter. Dynamical arrest like that of glasses is ubiquitous in nature. It occurs in a vast range of systems spanning microscopic to macroscopic scales. Despite its practical importance, a fundamental understanding of the glass transition is still lacking, making it one of the outstanding problems of condensed-matter science. For reviews see \cite{Binder2011,Ediger1996,Cavagna2009,Berthier2011,Biroli2013}.

Figure 1 illustrates the glass transition problem.  The central physical ingredient necessary for glassy slowing down is that of excluded volume interactions at high densities. Under such conditions motion is severely restricted through steric constraints, cf.\ Fig.\ 1(a), and particles can only move if the neighbouring particles that are blocking their path move before.  The higher the density the more collective motion becomes.  Cooperative relaxation leads to separation of timescales, where short scale motion (e.g. harmonic and anharmonic vibrations) is fast but the larger scale motion required for structural relaxation is slow. This becomes manifest in time correlations displaying metastability and decaying stretched exponential manner in time, indicative of a wide distribution of relaxation timescales, cf.\ Fig.\ 1(b). The typical relaxation time of supercooled liquids (inferred either from dielectric relaxation or from their viscosity) grows dramatically with decreasing temperature, a phenomenon which is quasi-universal, cf. Fig.1(c). By convention, when relaxation time becomes 100s (corresponding to a viscosity of fifteen orders of magnitude higher than that of a normal liquid), a liquid cannot be experimentally distinguished from a solid and such materials undergo a so-called experimental glass transition, corresponding to a falling out-of-equilibrium into the solid amorphous glass. 

A hallmark of dynamics close to the glass transition is dynamical heterogeneity, illustrated in Fig. 1(d) which shows the spatial pattern of relaxation in a two-dimensional supercooled Lennard-Jones mixture.  The slower the relaxation the more relaxation fluctuates in time and space: dynamics of systems close to the glass transition is fluctuation dominated and far from mean-field. It is important to note that the generic physics of the glass transition occurs in classical many-body systems in the absence of disorder, making them different from disordered systems such as spin-glasses.

\begin{figure}[t]
	\includegraphics[width=\columnwidth]{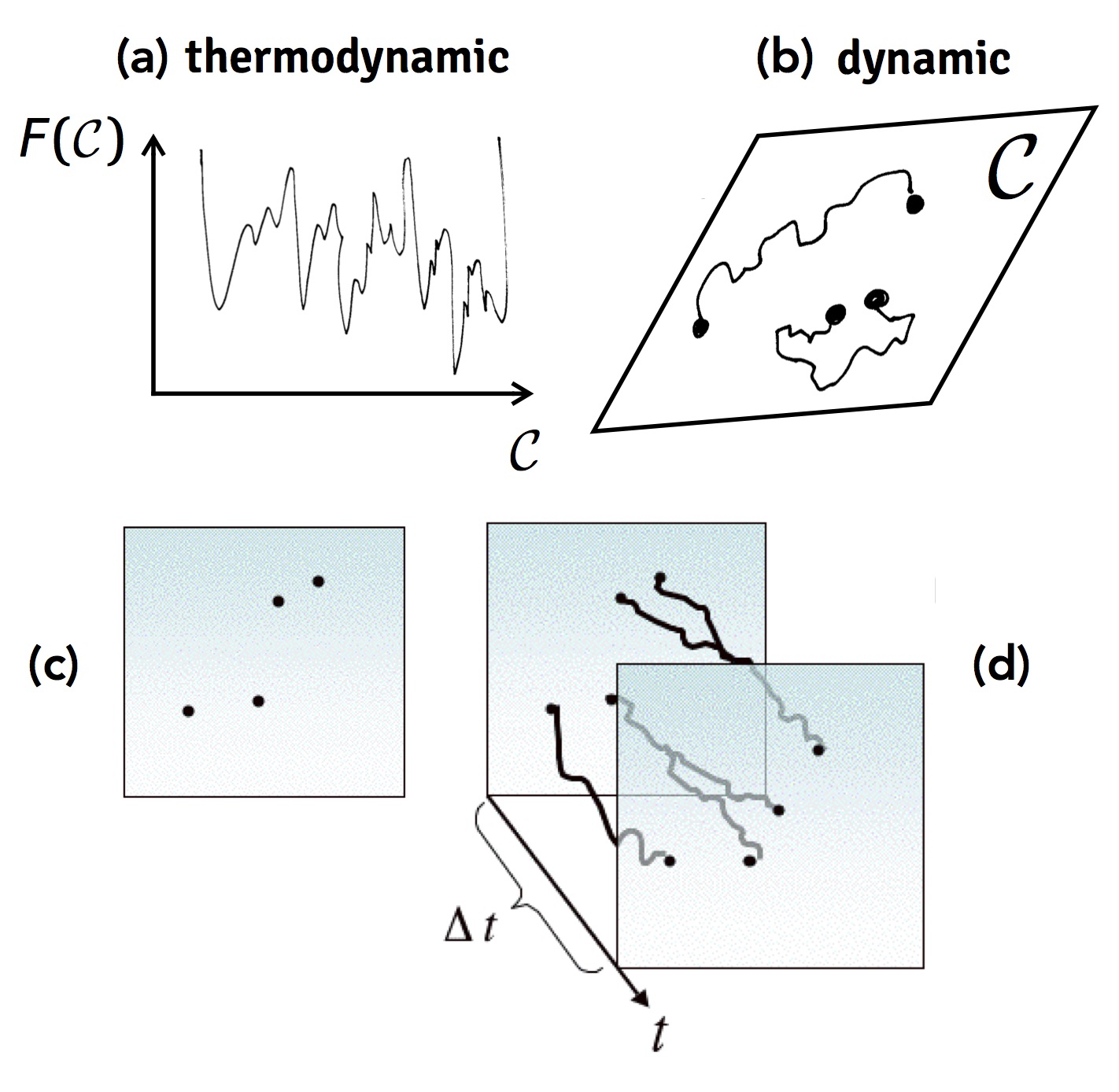}
\caption{
{\bf Theoretical perspectives on the glass transition:}
(a) ``Rugged'' free-energy of a glassy system as an illustration of thermodynamic perspectives such as RFOT. (b) Complex pathways in the
configuration space as an illustration of dynamic perspectives such as DF. (c) Sketch of localised effective excitations in space, and thus simple thermodynamics. (d) Trajectories of the dynamics can be complex due to constrained/facilitated propagation of excitations.
}
\label{fig2}
\end{figure}

\subsection{Theoretical perspectives on glasses}

A theoretical understanding of the slowdown and arrest of glass formers is a long-sought after goal of condensed matter science. The key distinguishing feature of glasses in contrast to more conventional condensed matter is that their observed dynamics is not accompanied by obvious structural changes.  Given that slowdown and arrest is widespread in soft materials it is expected that the underlying theory for such phenomena should be generic. 

Loosely speaking there are two classes of theoretical approaches.  One can be termed {\em thermodynamic} in the sense that it purports that the observed dynamics is a manifestation of underlying thermodynamic singularities. The most prominent of these is the so-called random fist-order transition (RFOT) theory \cite{Lubchenko2007,Parisi2010,Charbonneau2017}.  Figure 2(a) serves as a cartoon. It sketches the energy (or free-energy) of a glassy system as a function of its (perhaps coarse-grained) configuration.  Due to (effective) frustration this free-energy landscape in configuration space is ``rugged'' with a potential proliferation of low lying states that under the right conditions may dominate the thermodynamics, leading eventually to a static singularity at some non-zero temperature \cite{Lubchenko2007,Parisi2010,Charbonneau2017}. While this thermodynamic transition to an ``ideal glass state'' may be in an unobservable range, RFOT purports that is has consequences in the observable dynamics, e.g., with growing timescales due to growing correlation lengths of the underlying 
``amorphous order'' of the glass phase, and relaxation times diverging.\changes{
Within RFOT the growth of the primary relaxation time $\tau$ with decreasing temperature is often assumed to follow a Vogel-Fulcher-Tammann (VFT) law, $\tau = \tau_0 e^{A/(T - T_c)}$ \cite{Cavagna2009}, where $T_c$ is the temperature at which times would diverge, often identified with the temperature at which the static transition to the ideal glass state would occur.} Further intricacy of the landscape is also predicted within RFOT to lead to subsequent singularities deep in the glass phase, such as the ``Gardner transition'' and consequent changes in the rigidity properties of the amorphous solid \cite{Charbonneau2017}. While many of the ideas behind RFOT trace back to systems with quenched disorder such as spin-glasses \cite{Mezard1987}, its application to structural glasses is validated by exact results in the mean-field (large dimensional) limit for both liquids and hard spheres \cite{Parisi2010,Charbonneau2017}. 

An alternative perspective is {\em dynamical}, in particular the so-called dynamical facilitation (DF) theory \cite{Chandler2010}, which is the one that I favour. Figure 2(b) serves as its corresponding cartoon: the idea is that it is not the (statistical) properties of configurations or states that is relevant, but that of the pathways between them. That is, what matters is the connectivity of configuration space, and under the right conditions (as in cold liquids or dense colloids) the scarcity of dynamical pathways give rise to glassy slowing down, without the need for any underlying static singularity. Figures 2(b) and (c) illustrate the basic mechanisms by which this combination of simple thermodynamics and complex dynamics can be achieved. Imagine that the elementary excitations that give rise to motion are effectively localised and are scarce at low temperatures (e.g. they are energetically costly). These could for example be the kind of localised defects that are prominent in for example tilings and coverings at high densities \cite{Garrahan2009}. In such cases, thermodynamics would be simple and non-singular (say that of a low density gas of such excitations), cf.\ Fig.\ 2(c). However, if their dynamical evolution is subject to local {\em dynamical constraints} (as for example excitations that cannot appear or disappear spontaneously, but can branch or coalesce) then the structure of their trajectories could be much more complex, cf.\ Fig.\ 2(d).

\begin{figure*}[t]
\begin{center}
	\includegraphics[width=1.9\columnwidth]{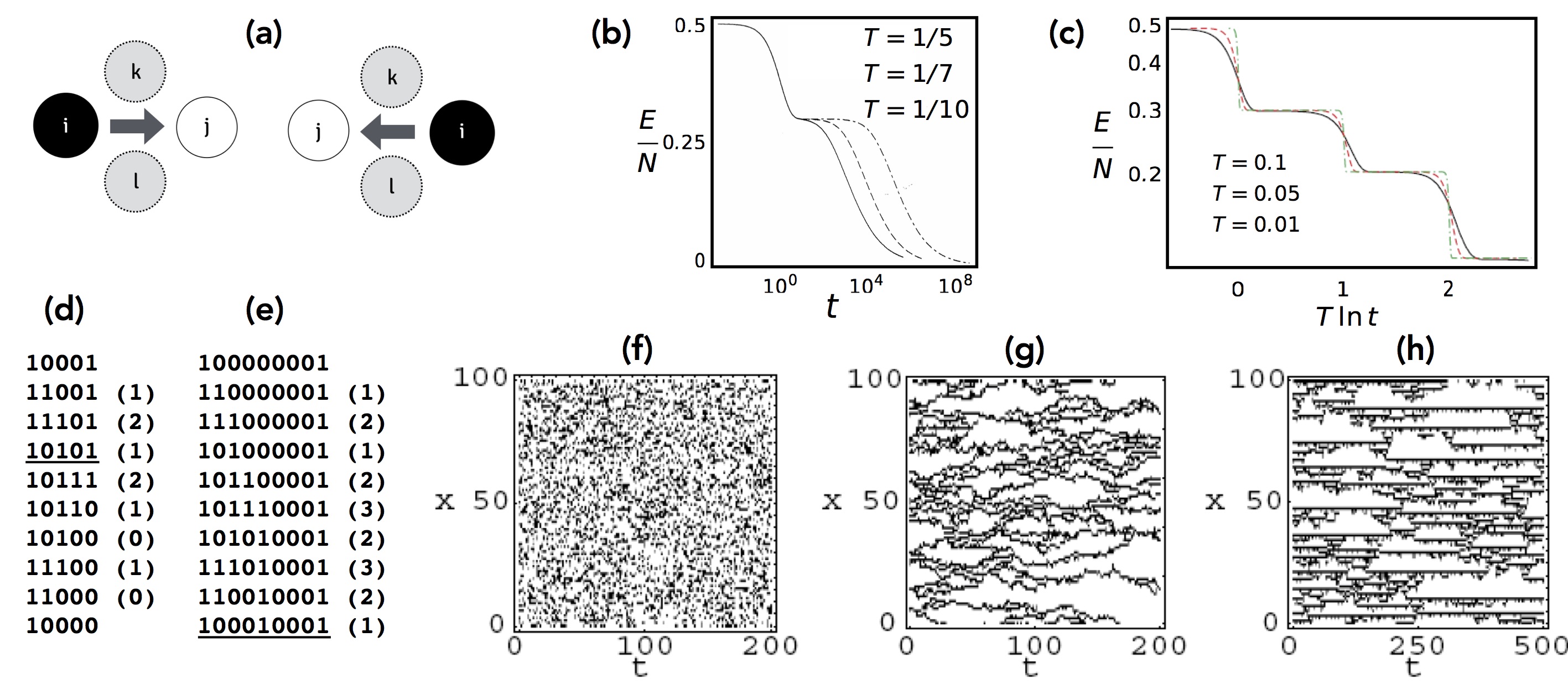}
\caption{
{\bf Slow cooperative dynamics due to constraints:}
(a) Sketch of a kinetic constraint.  (b) Relaxation of the energy in the FA model after a quench from $T=\infty$ to $T \ll J$.  
(c) Same for the East model, cf.\ Refs.\ \cite{Sollich1999,Ashton2005}.   
(d) Energetically optimal pathway for relaxation of a region of size $l=4$ in the East model. 
(e) Same for $l=8$. 
(f,g,h) Trajectories of the one-dimensional unconstrained, FA and East models, respectively, from Ref.\ \cite{Garrahan2002}.
}
\label{fig3}
\end{center}
\end{figure*}

\subsection{Kinetically constrained models of glasses}

The basic tenets of DF theory are realised explicitly in a class of idealised models known as kinetically constrained models (KCMs) \cite{Fredrickson1984,Jackle1991,Ritort2003,Garrahan2011} and whose study serves as the basis for DF insights \cite{Chandler2010}. The key idea is to encode the local effects of excluded volume interactions in the definition of the rates for transitions for the stochastic dynamics of these models. Figure 3(a) illustrates this: consider the transition where the particle labelled $i$ moves to occupy the empty space labelled $j$. Let us call the transition rate for this process $\gamma_{ij \to ji}$, and that for the reverse process $\gamma_{ji \to ij}$. In defining dynamics, we usually require that the rates obey detailed balance with respect to the Boltzmann distribution corresponding to some energy function $E$, such that $\gamma_{ij \to ji} / \gamma_{ji \to ij} = \exp \left(-\beta \Delta E_{ij}\right)$, where $\Delta E_{ij}$ is the difference in energies before and after the transitions. Assuming ergodicity in the dynamical rules, this should guarantee that this dynamics converges to the thermodynamics associated with this Boltzmann distribution. 

Now, dynamics-to-thermodynamics is many-to-one, that is, there are actually many ways to satisfy the same detailed balance condition with different definitions for the dynamical rules. This fact is often exploited in Monte Carlo simulations by defining an artificial dynamics that speeds up convergence to equilibrium (and where actual dynamical aspects are not of interest).  But it can also be exploited to model the actual evolution of systems with cooperative dynamics such as glasses.  In Fig. 3(a) the neighbouring particles $k$ and $l$ are not changing positions in either the forward or backward transitions, but their presence may affect the rates at which the transitions take place.  For example, if we think of excluded volume, the location of such neighbouring particles may determine whether the transitions can take place at all.  That is, we can make the rates dependent on these neighbours, 
\begin{equation}
\gamma_{ij \to ji} \to F(kl) \gamma_{ij \to ji}
\,\, , \,\,
\gamma_{ji \to ij} \to F(kl) \gamma_{ji \to ij}, 
\label{rates}
\end{equation}
and detailed balance is still obeyed,
\begin{equation}
\frac{F(kl) \gamma_{ij \to ji}}{F(kl) \gamma_{ji \to kl}}
=
\frac{\gamma_{ij \to ji}}{\gamma_{ji \to kl}}
=
e^{-\beta \Delta E_{ij}} .
\label{DB}
\end{equation}
In general, total transition rates are composed of an ``asymmetric part'', $\gamma_{ij \to ji}$ (in the sense that it changes depending on the direction of the transition), and a ``symmetric'' part which is the same in both directions - this latter we call a {\em kinetic constraint} as it does not affect the eventual stationary state, cf.\ \er{DB}, but does determine the actual dynamical evolution \cite{Baiesi2009}.

\subsubsection{Fredrickson-Andersen model}

Perhaps the simplest model that exploits the above idea to obtain non trivial relaxation dynamics is the one-spin facilitated Fredrickson-Andersen (FA) model \cite{Fredrickson1984,Ritort2003}. We will consider it in one dimension for simplicity. The FA model is defined in terms of binary variables $n_i = 0,1$ occupying the sites of a one dimensional lattice.  Its energy function is defined as, 
\begin{equation}
E = J \sum_{i=1}^N n_i, 
\label{E}
\end{equation}
leading to trivial thermodynamics (setting $k_{\rm B} = 1$), 
\begin{align}
Z  & = \left( 1 + e^{-J/T} \right)^N , \nonumber\\
c & \equiv \langle n_i \rangle = \frac{e^{-J/T}}{1 + e^{-J/T}} , \label{thd} \\
\langle n_i n_j \rangle &= \langle n_i \rangle \langle n_j \rangle = c^2
\;\; (i \neq j) \nonumber,
\end{align}
with no correlation between different sites and where the only relevant thermodynamic observable is the concentration $c$ of excited spins (i.e., those with $n=1$) in equilibrium. Of course, due to the absence of interactions in the energetics, thermodynamic properties are smooth all the way to $T=0$. 

The dynamics of the FA model is defined in terms of single spin-flip dynamics with the following rules,
\begin{align}
11 &\to 10 & {\rm rate} = 1-c \nonumber\\
10 &\to 11 & {\rm rate} = c \nonumber\\
11 &\to 01 & {\rm rate} = 1-c \nonumber\\
01 &\to 11 & {\rm rate} = c \label{FA} \\
010 &\to 000 & {\rm rate} = 0 \nonumber\\
000 &\to 010 & {\rm rate} = 0 \nonumber
\end{align}
The rates above are reversible and obey detailed balance with respect to $E$. The kinetic constraint is such that only sites who have an excited nearest neighbour can flip, and in that case they do so with the natural rates determined by the change in energy. But sites whose two nearest neighbours are unexcited cannot flip, even if energetically favourable.  The need for a neighbouring excitation to allow motion is often called {\em facilitation} \cite{Fredrickson1984,Ritort2003,Garrahan2002}. 

At high temperatures the kinetic constraint does not play a major role: in a typical equilibrium configuration at $T > J$ there are many excitations and most sites are therefore facilitated.  In contrast, for $T < J$ there is a conflict between the low number of excitations, $c \ll 1$, \er{thd}, in equilibrium configurations, and the need for them to be present to allow dynamics to proceed, \er{FA}.  Figure 3(b) illustrates this conflict: it shows the evolution of the energy per site after a ``quench'' from $T=\infty$ and $c=1/2$ to various low temperatures, $T < J$, so that the target state for the dynamics has $c \ll 1$.  Initially, relaxation is fast and $T$ independent as there are many excitations that facilitate dynamics - this evolution is similar to that of an unconstrained set of non-interacting spins. However, at some point most remaining excited sites that need to be relaxed are isolated: a plateau develops indicative of both a change of mechanism and a separation of timescales in the dynamics, and subsequent evolution is much slower and strongly $T$ dependent. 

This slow evolution is due to an effective {\em activated diffusion} of excitations.  
At low temperatures excitations are rare and typically an excited site is surrounded by a region of unexcited sites, $\ldots 00100 \ldots$.  While the central excitation here cannot relax, it can facilitate the excitation of one of its neighbours, say to the right, $\ldots 00110 \ldots$, a process that has rate $c$ to occur as it increases the energy by one unit of $J$.  Subsequently, the new excitation in turn can now facilitate the central spin to flip, $\ldots 00010 \ldots$, with rate $(1-c)/2$, the half coming from the fact that it could have been the rightmost excitation that relaxes, going back to the initial configuration (other processes are also possible, such as facilitating a third excitation, but this is suppressed by $c$, which is small at low $T$). The central excitation has then made a hop to the right with a rate that goes as $c(1-c)$.  Since it could have equally hopped to the left, overall what we get is effective diffusion of isolated excitations with diffusion rate $D \sim c$ when $c \ll 1$.  The upshot of this is that the relaxation of a site that is at distance $l$ from its nearest excitation requires the diffusion of this excitation to it, and its relaxation time goes as $\tau(l) \sim D^{-1} l^2$.  
From here we can get the typical relaxation time in equilibrium, $\tau_\alpha$: since in equilibrium the typical length between excitations is $l_{\rm eq} = c^{-1}$, we get, 
\begin{equation}
\tau_\alpha = \tau(l_{\rm eq}) \sim D^{-1} l_{\rm eq}^2 = c^{-(1+2)} \approx e^{3 J/T} .
\label{tauFA}
\end{equation}
The typical relaxation time of the FA model thus grows fast with decreasing temperature.  It does so in an Arrhenius form, but it is important to note that this is not due to a single barrier of size $3J$ but due to a combination of the way that the diffusion rate scales with temperature and the dynamic exponent that enters into $\tau(l)$. The results above are a consequence of the fact that the dynamics of the FA model in one dimension is fluctuation dominated as a consequence of the kinetic constraint.

\subsubsection{East model}

A second simple KCM of interest is the East model \cite{Jackle1991,Ritort2003,Garrahan2002}. Like the FA model it is defined in terms of binary variables on a lattice with a non-interacting energy function $E = J \sum_{i=1}^N n_i$, leading to the same simple thermodynamics, \er{thd}.  The difference lies in the kinetic constraints which as slightly more restrictive than those of the FA model, with the only (also single spin-flip) transitions allowed being,
\begin{align}
11 &\to 10 & {\rm rate} = 1-c \nonumber\\
10 &\to 11 & {\rm rate} = c \label{East}
\end{align}
that is, up spins facilitate flip of neighbouring spins in the eastwards direction only; i.e., in the East model the third and fourth transitions in \er{FA} are not allowed.  

The stricter constraints, \er{East}, mean that the mechanism that gives rise to activated diffusion of excitations in the FA model is not available in the East, as it requires facilitation in both directions.  This gives rise to further separation of timescale and more cooperative relaxation.  This is apparent from the evolution of the energy after a quench to low temperatures, Fig.\ 3(c): in contrast to the FA model case, Fig.\ 3(b), the energy decays in stages, each plateau associated with a hierarchy of the dynamics corresponding to relaxation over a particular range of lengthscales.

The hierarchical nature of the dynamics is easy to understand in terms of the optimal pathways for relaxation. Just like in the case of the FA model, at low temperatures excitations in typical equilibrium configurations will be isolated. 
While an up spin that is next to another up spin can relax immediately, first line in \er{East}, when they are further away intermediate steps that increase the energy are required. When the distance is $l=2$, as in $\cdots 101 \cdots$, an extra excitation needs to be created, $101  \to  111  \to  110  \to  100 $. Here the leftmost excitation facilitated the excitation of the middle site, which then facilitated the relaxation of the rightmost one. The energy barrier that needed crossing was $\Delta E(l=2) = J$ corresponding to the difference between the initial energy and the most energetic intermediate state visited. 

Consider now distance $l=4$, as in $\cdots 10001 \cdots$.  An obvious route to relaxing the rightmost spin is to excite all the ones in between so that it can be facilitated, and then de-excite them back, $ 10001  \to  11001  \to  11101  \to  11111 \to 11110 \to 11100 \to 11000 \to 10000$. In this case $\Delta E = 3J$ and in general such as procedure will give an energy barrier that scales with the length $l$ that needs to be span. One can however do better: an optimally energetic path for $l=4$ is shown in Fig.\ 3(d), and only requires $\Delta E(l=4) = 2J$. Figure 3(e) shows a similar construct for $l=8$.  The key idea is that one only needs to create an excitation at distance $l/2$ and the subsequent relaxation is that of a domain of half the length \cite{Sollich1999,Ritort2003}.  This means that 
\begin{equation}
\Delta E(2l) = \Delta E(l) + J
\Rightarrow
\Delta E(l) = J \lceil {\log_2 l} \rceil \; ,
\label{EEast}
\end{equation}
and the barrier only grows logarithmically with length.  (In the expression above $\lceil \cdot \rceil$ means the ceiling function, so that $l=2^{k-1} + 1, \ldots, 2^k$ all have barriers $\Delta E = J k$.)

At low temperatures relaxation will be activated and dominated by the smallest barrier that needs to be crossed.  From this argument, the timescale $\tau(l)$ to relax a region of size $l$ should go as \cite{Sollich1999,Ritort2003}
\begin{equation}
\tau(l) \sim e^{\Delta E(l)/T} \approx l^{z(T)} \; , 
\label{tEast}
\end{equation}
with 
\begin{equation}
z(T) = \frac{J}{T \ln 2} \; . 
\label{zEast}
\end{equation}
This means that relaxation in the East model obeys a scaling relation, but the dynamical exponent $z$ increases with decreasing $T$, indicating that dynamics becomes ``stiffer'' the lower the temperature.  These energetic arguments would suggest that the typical relaxation time at equilibrium would be given by the energy barrier at the typical distance between excitations, 
$\tau(l_{\rm eq}) \approx l_{\rm eq}^{z(T)} \approx \exp \left(J^2/T^2 \ln 2 \right)$  \cite{Sollich1999,Ritort2003}. However, as the length gets longer one needs also to consider the multiplicity of paths, which may give an entropic contribution \cite{Chleboun2013}.  In fact, while for modest distances the energetic contribution dominates and the scaling obeys \era{tEast}{zEast}, for lengths comparable to $l_{\rm eq}$, at low $T$, the entropic contribution is significant, and the overall typical relaxation time has been shown to be  \cite{Chleboun2013}
\begin{equation}
\tau_\alpha \sim \exp \left( \frac{J^2}{2 \ln 2 \, T^2} \right) .
\label{tauEast}
\end{equation}
This shows that in contrast to the FA model, the East model relaxes in a super-Arrhenius way. Furthermore, note that while $\tau_\alpha$ grows very fast with decreasing temperature, it only diverges at $T=0$.

On the one hand, the above results are interesting conceptually.  They show that, rather straightforwardly, local kinetic constraints give rise to hierarchical dynamics, cf.\ \era{tEast}{zEast}, and super-Arrhenius relaxation, \er{tauEast}.  On the other, they are also useful in practice if one consider whether a relaxation law such as the one of the East model (or its higher dimensional generalisations \cite{Garrahan2003,Berthier2003,Ashton2005,Chleboun2014}) is at all applicable to the observed phenomenology of supercooled liquids.  Figure 1(c) shows this indeed to be the case. It plots the relaxation time (or the viscosity) for a large number of supercooled liquids, fitting the data to the form $\ln \tau_\alpha \propto \left(J/T - J/T_{\rm o} \right)^2$, the so-called ``parabolic law'' \cite{Elmatad2009}, whose low $T$ behaviour scales with $T$ precisely like the East model, \er{tauEast}.  Below an ``onset'' temperature $T_{\rm o}$ dynamics becomes heterogeneous, and $T < T_{\rm o}$ is the regime where DF ideas become applicable.  In this regime the fit to the parabolic law is excellent, cf.\ Fig.\ 1(c). One can argue \cite{Elmatad2009,Elmatad2010} furthermore that the parabolic law provides a superior fit to the data to that found using the traditional VFT 
form \cite{Binder2011,Ediger1996,Cavagna2009,Berthier2011,Biroli2013}, $\ln \tau_{\rm VFT} \propto \left(T - T_K \right)^{-1}$, without the need to invoke an essential singularity at an unobservable $T_K > 0$.

The effect of the kinetic constraints on the dynamics becomes apparent if one looks at trajectories \cite{Garrahan2002}. \changes{Figures 3(f-h) show} one-dimensional equilibrium trajectories for three models for comparison: panel (f) corresponds to a spin model with energy function \er{E} and where dynamics is unconstrained, so that single spin-flip transitions $1 \to 0$ and $0 \to 1$ can occur with rates $1-c$ and $c$, respectively; panel (g) is the FA model; and panel (h), the East model.  While the three models have the same equilibrium distribution, their trajectories are very different.  In particular, both the FA and East models display pronounced space and time fluctuations (that give rise to heterogeneous dynamics) indicative of strongly correlated dynamics, thus realising the basic idea illustrated in Figs.\ 2(c,d).

\section{Large deviations and thermodynamics of trajectories}
\label{Lec2}

\subsection{Fluctuations in trajectory space}

The key insight of the DF approach \cite{Chandler2010} to glasses is that the interesting structure is to be found in the trajectories of the dynamics rather than in configurations, cf.\ Figs.\ 3(f-h). Take the case of the East model. Each time slice of an equilibrium trajectory such as that of Fig.\ 3(h) - see also Inset to Fig.\ 4(a) - corresponds to an equilibrium configuration.  As such, one-time (i.e., static) observables, like for example the magnetisation, have Gaussian distributions in the large size limit, given the non-interacting nature of the thermodynamics of the East model, cf.\ \era{E}{thd}.  Trajectories, however, have non-trivial correlations, something that is evident by the large ``space-time bubbles'' \cite{Garrahan2002}, cf.\ Fig.\ 3(h).  This means that in contrast to static observables, the distributions of {\em dynamical observables}, corresponding to time-integrated quantities such as for example the time integral of the magnetisation, $K_t = \int_{0}^{t} dt' M(t')$, must have non-Gaussian distributions \cite{Merolle2005}.  This is indeed the case, see Fig.\ 4(a): $P(K_t)$ has pronounced non-Gaussian tails in the {\em large deviation} regime (i.e., away from the mean and variance). The reason that time-integrated observables reveal information that static observables do not is that the moments of quantities like $K_t$ contain time-integrals of time-correlation functions of its integrand, and thus encode both space and time correlations.  Furthermore, a distribution like that of Fig.\ 4(a) is suggestive of an interesting phase behaviour in trajectory space, whose study requires a method for ensembles of trajectories that is analogous to the more standard equilibrium ensemble methods of equilibrium statistical mechanics.  This can be readily defined using the machinery of large deviations \cite{Touchette2009}.

\begin{figure*}[t]
\begin{center}
	\includegraphics[width=1.9\columnwidth]{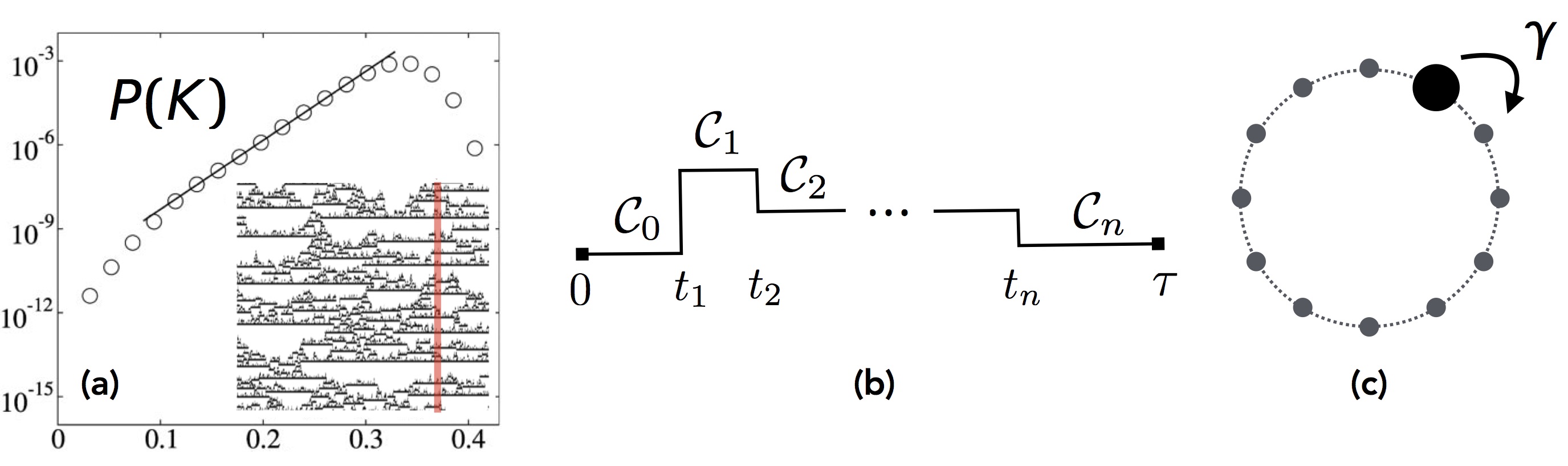}
\caption{
(a) Probability of the activity in the East model, cf.\ \cite{Merolle2005}.
(b) Sketch of a trajectory in a continuous time Markov chain.
(c) Particle hopping unidirectionally in a one-dimensional ring.
}
\label{fig4}
\end{center}
\end{figure*}

\subsection{Stochastic dynamics}

In what follows for concreteness I will consider stochastic dynamics corresponding to continuous time Markov chains.  Consider a classical stochastic system evolving as a continuous time Markov chain.  The Master Equation (ME) for the probability reads~\cite{Gardiner2004}
\begin{align}
\partial_{t} P(C,t) = \sum_{C' \neq C} W({C' \to C}) P(C',t) 
\nonumber \\
- R(C) P(C,t) , 
\label{ME1}
\end{align}
where $P(C,t)$ indicates the probability of the system being in configuration $C$ at time $t$, 
$W({C' \to C})$ is the transition rate from $C'$ to $C$, and $R({C}) = \sum_{C' \neq C} W({C \to C'})$ the escape rate from $C$. The ME can be written in operator form,  
\begin{equation}
\partial_{t} |P(t) \rangle = \W |P(t) \rangle ,
\label{ME2}
\end{equation}
with probability vector $|P(t)\rangle$ 
\begin{equation}
|P(t)\rangle = \sum_{C} P(C,t) | C \rangle
\label{P} ,
\end{equation}
where $\{ | C \rangle \}$ is an orthonormal configuration basis, $\langle C | C' \rangle = \delta_{C,C'}$.  The Master operator $\W$ is defined as,
\begin{align}
\W &= \sum_{C,C' \neq C} W({C \to C'}) |C' \rangle \langle C|
- \sum_{C} R({C}) |C \rangle \langle C|
\label{W}
\end{align}
The stochastic generator $\W$ is in general non-Hermitian but has a real spectrum, with its largest eigenvalue equal to zero. The associated right eigenvector corresponds to the probability vector of the stationary state, 
\begin{equation}
\W | {\rm ss} \rangle = 0 , \label{ss}
\end{equation}
while the corresponding left eigenvector 
is the ``flat'' or ``trace'' state
\begin{equation}
\fs = \sum_{C} \langle C| , \label{flat}
\end{equation}
and $\fs \W = 0$ is the statement of probability conservation. 

The dynamics described by the ME is realised by stochastic trajectories.  Each trajectory is a particular realisation of the noise that gives rise to a time record of configurations and of waiting times for jumps between them, observed up to some \changes{time}.  Figure 4(b) sketches such trajectories: the trajectory starts in configuration $C_{0}$ and then there is a succession of configuration changes,  
$C_{0} \to C_{1} \to C_{2} \to \ldots \to C_{n}$ occurring at times $t_1, t_2, \ldots, t_n$. Between the last jump and the end of the trajectory at $\tau$ there is no further change in configuration.  The evolution described by the ME is recovered when averaging over all stochastic trajectories.

\subsection{Dynamical large deviations}

\subsubsection{Example}

To introduce the ideas, Let us consider first an elementary example which is exactly solvable.  Consider unidirectional hopping of a particle with rate $\gamma$ between nearest neighbouring sites of a one-dimensional ring, as in Fig.\ 4(c). The possible positions of the particle \changes{determine} the set of configurations,  $\{ | C \rangle \} = \{ | x \rangle :  \; x=1,\ldots,L \}$.  The Markov generator $\W$ reads,
\begin{equation}
\W = \gamma \sum_{x=1}^{L} |x+1\rangle \langle x| - \gamma \sum_{x=1}^{L} |x \rangle \langle x| .
\label{Wex}
\end{equation}
If we denote a stochastic trajectory of total time $t$ by $\omega_t$, the simplest dynamical observable one can consider is the total number of jumps in the trajectory $K(\omega_t)$.  This observable for example can be used to classify trajectories according to how dynamically ``active'' they are: a more active trajectory will be one with a higher value of $K$, and a less active, one with a lower value of $K$.  The distribution of $K$, 
\begin{equation}
P_{t}(K) = \sum_{\omega_t} {\rm Prob(\omega_t)} \delta \left[ K(\omega_t) - K \right] , \label{PK}
\end{equation}
where ${\rm Prob(\omega_t)}$ indicates the probability of getting trajectory $\omega_t$ with the dynamics \er{Wex}, provides information about the statistical properties of the dynamics. For the simple example we are considering this is of course a Poisson distribution, 
\begin{equation}
P_{t}(K) = e^{-\gamma t} \frac{(\gamma t)^{K}}{K!} .
\label{PKex}
\end{equation}
Using the Stirling approximation for the factorial, one finds that at long times the probability acquires a {\em large deviation} form, becoming an exponential in time, times a function of the {\em intensive} observable $K/t$, 
\begin{equation}
\changes{
P_{t}(K) \sim 
e^{-t \left[ \gamma - \frac{K}{t} + \frac{K}{\gamma t} \ln \left(\frac{K}{\gamma t} \right) 
\right]} \; .
}
\label{PKexLD}
\end{equation}
Essentially the same information as in the probability is contained in the moment generating function (MGF), 
\begin{equation}
Z_t(s) = \sum_K P_t(K) e^{-s K} \; ,
\label{Zs}
\end{equation}
which for our simple problem reads, 
\begin{equation}
Z_t(s) = e^{t \gamma \left( e^{-s} - 1 \right) } \; .
\label{Zsex}
\end{equation}
We see that \er{Zsex} also has a large deviation form of an exponential in time, times a function of the conjugate variable $s$ to the observable $K$.

\subsubsection{Generalisation}

The above approach directly generalises to any system with stochastic dynamics described by a ME such as \er{ME1}. A dynamical order parameter $K$ is a time-extensive function of the trajectory. In its most general form it is reads,
\begin{equation}
K(\omega_t) = t \sum_{C,C' \neq C} q_{C,C'}(\omega_t) \alpha_{C \to C'} + 
t\sum_C \mu_C(\omega_t) \beta_C .
\label{Kgen}
\end{equation}
This form indicates that dynamical observables can increase either when the trajectories make jumps, or by accumulating the value of a static observable between jumps. The first summation in \er{Kgen} corresponds to advancing $K$ by $\alpha_{C \to C'}$ for each jump in the trajectory between configurations $C$ and $C'$. The total number of jumps in trajectory $\omega_t$ between $C$ and $C'$ is indicated by $t q_{C,C'}(\omega_t)$, with $q$ sometimes called the dynamical or empirical {\em flux} \cite{Touchette2009}. The second summation corresponds to the time-integral of a static observable of the configurations. The total time spent in configuration $C$ in trajectory $\omega_t$ is indicated by $t \mu_{C}(\omega_t)$, with $\mu$ referred to as the {\em empirical measure} (as it gives the estimate of the distribution over configurations that can be inferred from that specific trajectory).  For example, for the case of the total number of jumps of the example above we would set all $\alpha_{C \to C'}=1$ and all $\beta_C=0$.  Conversely, if we were time-integrating say a magnetisation we would set all $\alpha_{C \to C'}=0$ and $\beta_C$ equal to the magnetisation of the configuration, $\beta_C = M(C)$.

The distribution \er{PK} for a dynamical order parameter such as \er{Kgen} in general acquires a LD form at long times \cite{Touchette2009,Lecomte2007,Garrahan2009}, 
\begin{equation}
P_{t}(K) \sim 
e^{-t \varphi(K/t)} \; ,
\label{PKLD}
\end{equation}
where $\varphi(k)$ is called the LD {\em rate function}.  Similarly the MGF, \er{Zs}, also acquires a LD form \cite{Touchette2009,Lecomte2007,Garrahan2009}, 
\begin{equation}
Z_t(s) \sim e^{t \theta(s) } \; , 
\label{ZsLD}
\end{equation}
with $\theta(s)$ called alternatively the {\em scaled cumulant generating function} (SCGF) (as its derivatives evaluated at $s=0$ give the cumulants of $K$ scaled by time), or simply the {\em LD function}. The rate function and the SCGF are related by a Legendre transform, which for our choice of convention reads \cite{Touchette2009,Lecomte2007,Garrahan2009}, 
\begin{equation}
\theta(s) = - \min_k \left[ k s + \varphi(k) \right] .
\label{LT}
\end{equation}

This approach provides a direct generalisation of the ensemble method of equilibrium statistical mechanics.  The analogy is summarised in Table I. In equilibrium the relevant ensemble is that of configurations $C$, while for dynamics it is that of trajectories $\omega$.  In dynamics, the large size limit is that of long times. Static order parameters are functions extensive in volume, such as the magnetisation $M$ of a configuration in a magnetic problem; the corresponding intensive order parameter is obtained by scaling out the system size, $m = M/V$.  For dynamics, order parameters are functions of the trajectories extensive in time, such as time-integrated quantities $K$; the corresponding intensive order parameters are obtained as $k = K/t$. At large volumes (resp.\ long times) the distribution of the order parameter has a LD form, cf.\ \er{PKLD}, and is determined by an entropy density, i.e., the log of the number of configurations (resp.\ trajectories) where the order parameter takes a certain value, scaled by size.  In the case of the magnetisation, for the static problem this is the Gibbs free-energy; in the case of dynamics, it is the rate function, \er{PKLD}.  Associated to each order parameter there is a conjugate field.  In the static case, for example when the order parameter is the magnetisation $m$, the conjugate field is the magnetic field. For dynamics, the conjugate field to a dynamical observable $K$ is the {\em counting field} $s$, which is also the variable of its MGF. At large size (long time) the MGF has a LD form, cd.\ \er{ZsLD}, and is determined by a free-energy, in the static case, in the magnetic language we are using, it corresponds to the Helmholtz free-energy; for dynamics, it is the SCGF, \er{ZsLD}.

\begin{table*}
\begin{center}
\begin{tabular}{rclcl}
\hline \hline \\
 && {\bf statics} && {\bf dynamics} \\
\\ \hline \\
ensemble: && configurations $C$ && trajectories $\omega$ \\
\\ \hline \\
 thermodynamic limit: && $V \to \infty$ && $t \to \infty$ \\
\\ \hline \\
 order parameter: && $m = M/V$ && $k = K/t$ \\
\\ \hline \\
 entropy density: && $G(m)/V$ (Gibbs free-energy) && $\varphi(k)$ (rate function) \\
\\ \hline \\
conjugate field: && $h$ && $s$ (counting field)\\
\\ \hline \\
free-energy: && $F(h)/V$ (Helmholtz free-energy) && $\theta(s)$ (SCGF) \\
 \\ \hline
 \hline
\end{tabular}
\end{center}
\caption{Comparison between equilibrium and trajectory ensembles.}
\end{table*}

The SCGF $\theta(s)$ is the free-energy for trajectory ensembles, so in analogy with equilibrium, its analytic structure tells us about the phase structure of the dynamics. Computing free-energies is often hard, but there is one property that can be exploited to obtain $\theta(s)$.  In dynamics time is a special direction, which in turn allows to calculate the MGF (which plays the role of a partition sum) in terms of a ``transfer matrix''.  This reduces the problem of calculating a partition sum, to the often simpler problem of maximising an operator.  In particular, the MGF, \er{Zs}, can be written as \cite{Touchette2009,Lecomte2007,Garrahan2009}
\begin{equation}
Z_t(s) = \fs e^{t \W_s} | P_0 \rangle , \label{Zs2}
\end{equation}
where $P_0$ is the initial distribution of configurations.  The operator $\W_s$ is obtained by deforming or {\em tilting} the Markov generator $\W$.  For a general dynamical observable, \er{Kgen}, the tilted generator reads \cite{Touchette2009,Lecomte2007,Garrahan2009}
\begin{align}
\W_s = & \sum_{C,C' \neq C} e^{-s \, \alpha_{C \to C'}} W({C \to C'}) |C' \rangle \langle C| \nonumber \\
& ~~~~~~~~~~ - \sum_{C} \left[ R({C}) + s \, 
\changes{
\beta_C
}
 \right] |C \rangle \langle C| .
\label{Ws}
\end{align}
The SCGF is the largest eigenvalue of $\W_s$, so that at long times from \er{Zs2} we recover \er{ZsLD}.

\begin{figure*}[t]
\begin{center}
	\includegraphics[width=1.9\columnwidth]{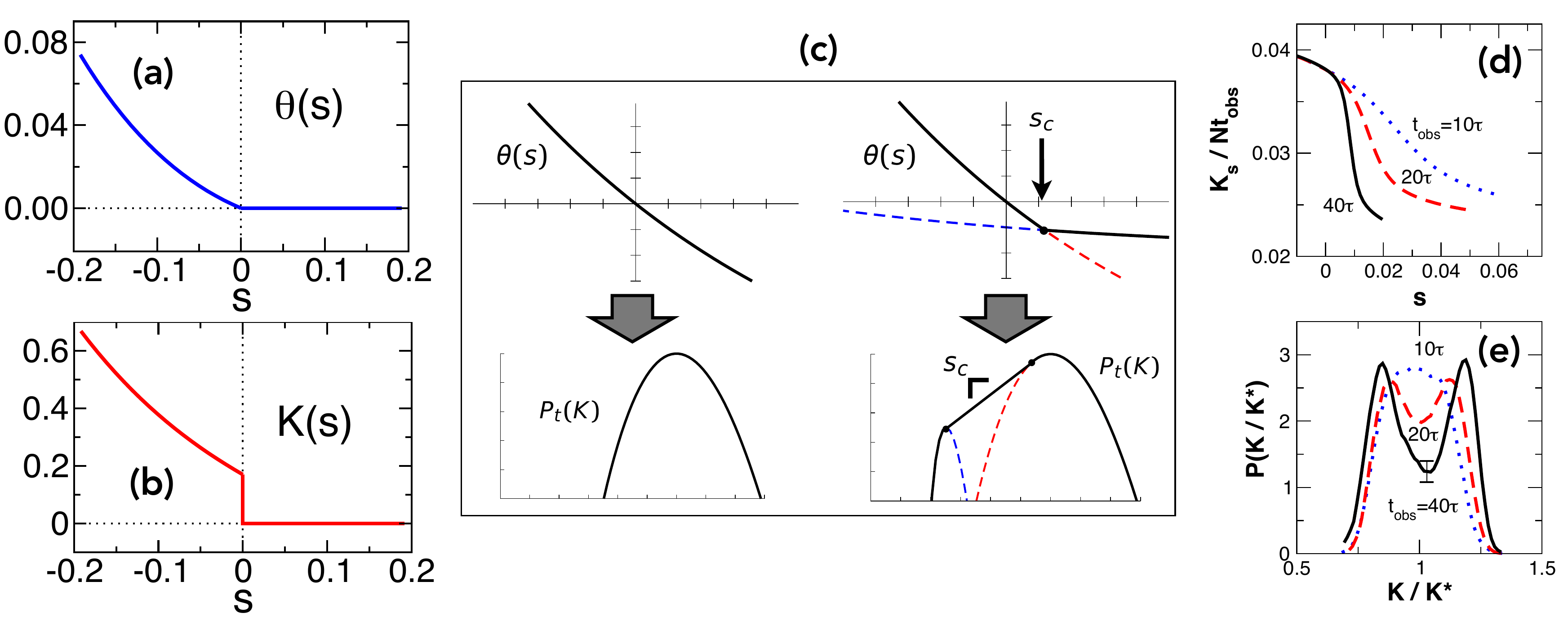}
\caption{
{\bf Phase transitions in trajectory space:}
(a) Mean-field approximation to the SCGF of the East model showing a singular change at $s=0$. 
(b) The corresponding order parameter $\langle K \rangle_s$ showing a discontinuous jump at $s=0$ indicative of a phase transition between dynamical active and inactive phases. Cf.\ Refs.\ \cite{Garrahan2007,Garrahan2009}.
(c) Connection between SCGF or free-energy and  
\changes{the logarithm of the}
order parameter distribution
via Legendre transform. The rightmost panels correspond the the case of a first-order phase transition. 
(d) Same as (b) but for an atomistic model of a supercooled liquid. (e) Order parameter distribution at coexistence: bimodality indicates a first-order dynamical transition in the supercooled liquid in analogy with KCMs. From Ref.\ \cite{Hedges2009}. 
}
\label{fig5}
\end{center}
\end{figure*}

\subsection{Phase transitions in trajectory space}

Let us come back to the problem of the East model. It was mentioned above that the large dynamical fluctuations evident in the East model trajectories, cf.\ Fig.\ 4(a), suggested a non trivial phase behaviour of the dynamics.  We will now see how with the LD methods described above it is possible to show that there is indeed a novel class of dynamical phase transitions in KCMs such as the East model \cite{Garrahan2007}.

The Markov generator for the East model dynamics is given by \cite{Jack2006}
\begin{equation}
\W = \sum_{i} 
n_{i-1}
\left[ \epsilon \sigma_{i}^{+} + \sigma_i^- - \epsilon (1-n_i) - n_i \right] ,
\label{WEast}
\end{equation}
where we have scaled out a factor of $(1-c)$ and defined $\epsilon = c/(1-c)$. The factors in square brackets in \er{WEast} correspond to flipping the spins and their associated escape rates, and the factors of $n_i$ before the brackets are the kinetic constraints. 

Let us consider as trajectory observable the {\em dynamical activity} \cite{Garrahan2007,Lecomte2007,Garrahan2009} (sometimes also termed frenesi \cite{Baiesi2009}) defined as the total number of configuration changes in a trajectory.  In the general definition of \er{Kgen} this would correspond to,
\begin{equation}
K(\omega_t) = t \sum_{C,C' \neq C} q_{C,C'}(\omega_t) ,
\label{act}
\end{equation}
i.e., all jumps between configurations are counted equally.  This is the natural order parameter for ``glassiness'', as it measures whether trajectories display a lot of motion (the kind of dynamics one would associate with liquid-like ergodic relaxation) or very little motion (what one would associate with glassy slow down and arrest), and it does so in an ``order agnostic'' manner in the sense that it makes no assumptions as to which configurations give rise to which kind of dynamics. 

The tilted generator deformed according to \er{act} is then, cf.\ \er{Ws}, 
\begin{equation}
\W_s = \sum_{i} 
n_{i-1}
\left[ e^{-s} \left( \epsilon \sigma_{i}^{+} + \sigma_i^- \right) - \epsilon (1-n_i) - n_i \right] ,
\label{WEasts}
\end{equation}
where $\sigma_{i}^{\pm}$ are Pauli raising and lowering operators acting on site $i$, and $n_{i} = \sigma_{i}^{+} \sigma_{i}^{-}$. Calculating the largest eigenvalue of the operator \er{WEasts} for arbitrary $s$ is a difficult challenge, so we will employ a series of approximations to simplify the problem so as to get a result that - while not quantitatively accurate - gives a qualitative idea of what the LD approach reveals. 

We first approximate the spin problem described by \er{WEasts} where sites can only  have single occupation, by a similar ``bosonic'' problem \cite{Cardy1999,Jack2006} where the single occupation restriction is lifted. The tilted operator reads,
\begin{equation}
\W_s^{\rm bos} = \sum_{i} 
n_{i-1}
\left[ e^{-s} \left( \epsilon a_{i}^{\dagger} + a_i \right) - \epsilon - n_i \right] ,
\label{WEastsb}
\end{equation}
where now $n_i = a_{i}^{\dagger} a_i$. NB: we are using the Doi-Peliti formalism, see e.g.\ Ref.\ \cite{Cardy1999,Tauber2005,Garrahan2009,Jack2006}, where the representation of creation and annihilation operators is not the standard one of quantum mechanics, but one more convenient for stochastic dynamics,
\begin{equation}
a^\dagger | n \rangle = | n + 1 \rangle 
\; , \;\;\;
a | n \rangle = n | n - 1 \rangle 
\; , \;\;\;
[a, a^\dagger] = 1 \; .
\label{ada}
\end{equation}
The approximation \er{WEastsb} is good for $\epsilon \ll 1$, corresponding to low temperatures, as multiple site occupancy is suppressed; cf.\ reaction-diffusion problems \cite{Cardy1999,Tauber2005}.

The largest eigenvalue of \er{WEasts} can be estimated via a variational approximation, for example by maximising the expectation value of $\W_s^{\rm bos}$ in a coherent state basis, cf.\ Ref.\ \cite{Garrahan2009}.  In practice this amounts to solving the Euler-Lagrange equations, 
\begin{equation}
\frac{\partial \W_s^{\rm bos}}{\partial a_i} = \frac{\partial \W_s^{\rm bos}}{\partial a_i^\dagger} = 0 ,
\label{EL}
\end{equation}
where operators are treated as c-numbers and the generator is normal ordered. We make a further mean-field approximation and drop the site index to obtain the equations, 
\begin{equation}
\left\{
\begin{array}{rcl}
0= \frac{\partial \W_s^{\rm bos}}{\partial a} & = & a^\dagger 
\left[ e^{-s} \left( \epsilon a^{\dagger} + a \right) - \epsilon - n \right] 
+ n \left(e^{-s} - a^\dagger \right) \\
\\
0= \frac{\partial \W_s^{\rm bos}}{\partial a^\dagger} & = & a 
\left[ e^{-s} \left( \epsilon a^{\dagger} + a \right) - \epsilon - n \right] + n \left(e^{-s} \epsilon - a \right)  \\
\end{array}
\right.
\nonumber
\end{equation}
Under $a = \epsilon a^\dagger$ these two equations become equivalent, and there are two solutions,
\begin{equation}
a_{\rm I} = \frac{3}{4} \epsilon e^{-s} + \frac{\epsilon}{4} \sqrt{9 e^{-s} - 8} 
\;\;\; {\rm and} \;\;\; a_{\rm II} = 0 .
\label{aa}
\end{equation}
Inserting the two possible solutions of the EL equations back into $\W_s^{\rm bos}$ we get two variational estimates for the SCGF, $\theta_{\rm I}(s)$ which is dependent on $s$, and $\theta_{\rm II}(s) = 0$ for all $s$.  Of these two possible branches we have to choose the largest one for each value of $s$. Due to convexity of the SCGF, $\theta_{\rm I}(s)$ crosses from positive to negative at $s=0$, and therefore $\theta(s)=\theta_{\rm I}(s)$ for $s<0$ and $\theta(s)=\theta_{\rm II}(s)=0$ for $s>0$; see Fig.\ 5(a). The SCGF if singular at $s=0$ where the two solutions cross, corresponding to a {\em first order transition} in the ensemble of trajectories \cite{Garrahan2007,Garrahan2009}. This is evident in the behaviour of the activity $K$. The $s$-dependent average activity, 
\begin{equation}
\langle K \rangle_s = \frac{\sum_{K} P_t(K) K e^{-s K}}{Z_t(s)} ,
\label{Ks}
\end{equation}
is the average activity when the probability of a trajectory $\omega_t$ is tilted by $e^{-s K(\omega_t)}$, and thus corresponds to the activity of trajectories of dynamics away from the typical one (corresponding to $s=0$) - such reweighed ensemble of trajectories is sometimes called the $s$-ensemble \cite{Hedges2009}. As such, $\langle K \rangle_s$ serves as a dynamical order parameter for classifying the ensemble of trajectories controlled by $s$ (cf.\ the magnetisation and magnetic field in the static case, see Table I). 

At long times $\langle K \rangle_s$ is obtained from the first derivative of the SCGF with respect to $s$, 
\begin{equation}
k(s) = \lim_{t \to \infty} \frac{\langle K \rangle_s}{t} = - \theta'(s) .
\label{Ksinfty}
\end{equation}
As Figs.\ 5(a,b) show, our mean-field estimate of $\theta(s)$ for the East model has a first order singularity at $s=0$ and the corresponding order parameter shows a discontinuous jump at the transition point. 

Figure 5(c) provides a reminder of the connection between the free-energy and the order parameter distribution, as connected by the Legendre transform. When the free-energy is analytic, the distribution is unimodal. This is indicative of the system having a single phase. Instead when the free-energy is non-analytic, for example when two branches cross at some value of the controlling field, the corresponding distribution is one associated to two-phases made convex by a Maxwell construction. 
The same relation between SCGF and rate function holds for dynamics. In the case of the East model the transition at $s=0$, Figs.\ 5(a,b), is one between an {\em active phase} with finite activity for $s \leq 0$, and an {\em inactive phase} of vanishing activity for $s>0$ \cite{Garrahan2007,Garrahan2009}. 

The qualitative picture gleaned from the crude approximations above is confirmed by more quantitative studies of the East model and other KCMs \cite{Garrahan2007,Garrahan2009,Elmatad2010b}. Furthermore, it has been shown that similar active-inactive dynamical transitions are present in more realistic models of glasses \cite{Hedges2009,Pitard2011,Speck2012,Speck2012b}, cf.\ Fig.\ 5(d,e), and the fluctuations associated to such dynamical first-order transitions manifest in the heterogenous pattern of relaxation in these systems, cf.\ Fig.\ 1(d).

\section{Ergodicity and non-ergodicity in closed quantum systems}
\label{Lec3}

\subsection{Quantum equilibration} 

Generic quantum many-body systems - evolving unitarily according to their Hamiltonian - are said to {\em equilibrate}, meaning that at long times their state becomes indistinguishable, from the point of view of expectation values of observables, from the time-integrated state \cite{Reimann2008,Linden2009,Short2011,Short2012,Reimann2012}; for reviews see \cite{Gogolin2016,DAlessio2016,Borgonovi2016}. This occurs due to dephasing of the state of the system in the energy eigenbasis (a condition we assume to hold, but which can be somewhat relaxed). Describing the state of a system in terms of its density matrix, we have for its time evolution, 
\begin{equation}
\rho(t) = U_{t} \, \rho_0 \, U_{t}^\dagger ,
\label{rhot}
\end{equation}
where 
\begin{equation}
U_t = e^{- i t H} ,
\label{U}
\end{equation}
with $H$ the Hamiltonian of the system (and we have set $\hbar = 1$ from here onwards). Expanding in the energy eigenbasis we get
\begin{align}
\rho(t) = & \sum_n 
| n \rangle \langle n | \, \rho_{nn} \nonumber \\
& + 
\sum_{n \neq m} 
| n \rangle \langle m | \, \rho_{nm} \, e^{-i (E_m-E_n) t} \; .
\label{rhott}
\end{align}
Due to the absence of degeneracies in the energy differences, the oscillatory terms in the second line of \er{rhott} become negligible at large times, and in the long time limit the state becomes indistinguishable in practice from the time-integrated state 
\begin{equation}
\rho(t)
\longrightarrow
\;
\overline{\rho(t)} = 
\lim_{t \to \infty} 
\frac{1}{t} \int_0^t U_{t'} \rho U_{t'}^\dagger dt' , 
\label{rhottt}
\end{equation}
meaning specifically,  
\begin{equation}
\left| \langle A(t) \rangle - {\rm Tr}
\left[ A \,\overline{\rho(t)}
\right] 
 \right|
~~ \text{small} . 
\label{small}
\end{equation}

Figure 6(a) illustrates what in practice is understood as equilibration in a  quantum system.  It describes the evolution of a spin-$1/2$ chain with repulsive density-density interactions that decay with distance as $r^{-6}$,
\begin{equation}
H = \Omega\sum_{i=1}^L \sigma^{x}_i -\mu \sum_{i=1}^L n_i 
+ \frac{V}{2} \sum_{i\neq j}^L \frac{n_i n_j}{|m-k|^6} \; , 
\label{HRyd}
\end{equation}
\changes{where $n_i = \sigma^+_i \sigma^-_i$. \Er{HRyd} describes }
a problem of relevance to Rydberg atoms \cite{Lesanovsky2010}.  This Hamiltonian is generic, in the sense that it is not integrable and due to the nature of the interactions the likelihood of degenerate energy gaps is negligible, certainly in the large size limit.  Equilibration becomes evident if one considers the unitary dynamics generated by \er{HRyd}: Fig.\ 6(a) shows the behaviour of local observables (i.e., observables that correspond to sums of local terms) in a numerical simulation of a finite system (from direct diagonalisation); at long times expectation values become stationary and close to their time average, something that becomes more precise with increasing system size. (Of course, for a finite system like this one there are renewals eventually, but these occur on timescales much larger than the ones shown.)

It is important to note that ``equilibration'' as used in the context of quantum many-body systems is different from the usual meaning attributed to this concept in the case of classical stochastic systems where it is closer to the idea of ``thermalisation'' (see below).  That is, that a quantum system equilibrates does not mean it is also ergodic, only that observations in long time state are in practice stationary in the large size limit, cf.\ Fig.\ 6(a). This effectively stationary state can still depend on details of the initial conditions, as we will see below.

\begin{figure*}[t]
\begin{center}
	\includegraphics[width=1.9\columnwidth]{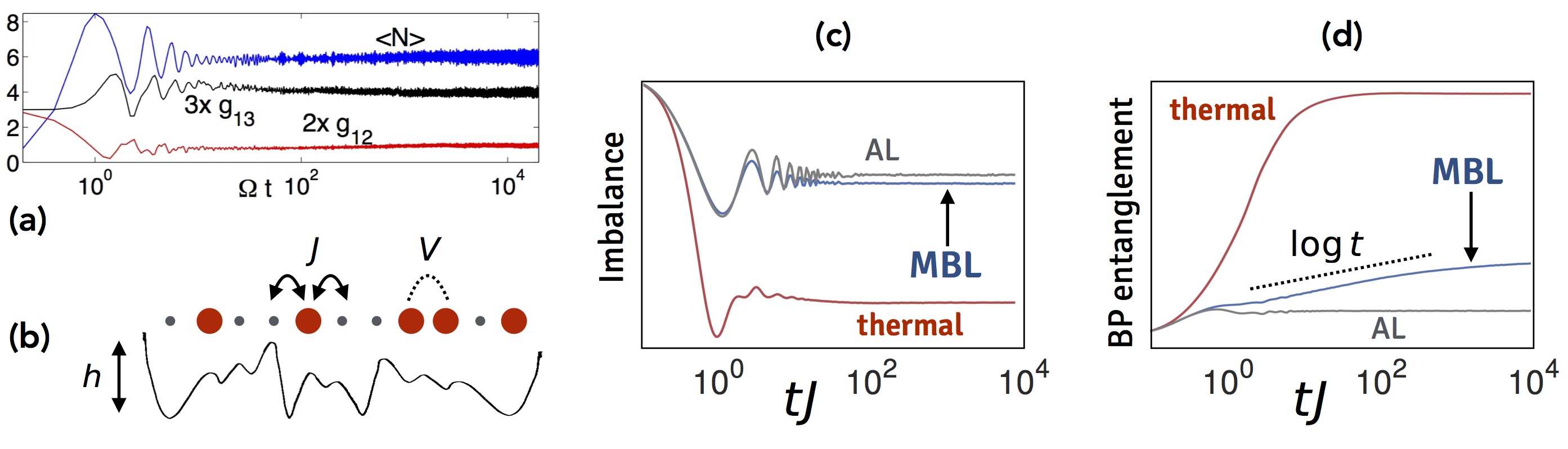}
\caption{
{\bf Quantum equilibration and many-body localisation:}
(a) Example of quantum equilibration in a spin chain modelling Rydberg atoms, from Ref.\ \cite{Lesanovsky2010}.
(b) Sketch of a typical MBL setup: spinless fermions with nearest neighbour hopping $J$, density interactions $V$ in the presence of a disordered local potential.
(c) Evolution of the imbalance under thermalisation (small $h$, red), MBL (large $h$, blue), and Anderson localisation ($V=0$, $h \neq 0$, grey) conditions.
(d) Corresponding evolution of the bipartite entanglement entropy, cf.\ \cite{Bardarson2012,Serbyn2013}.
}
\label{fig6}
\end{center}
\end{figure*}

\subsection{Quantum thermalisation}

Beyond equilibration, most quantum many-body systems are believed also to {\em thermalise} \cite{Gogolin2016,DAlessio2016,Borgonovi2016}.  The meaning is the following. Consider two partitions $A$ and $B$ of an overall closed many-body quantum system. Let us assume further for simplicity that the initial state is a pure state $| \psi_0 \rangle$, so that 
\begin{equation}
\rho_0 = | \psi_0 \rangle \langle  \psi_0 | \; .
\label{pure}
\end{equation}
It follows, that at all subsequent times the state of the system is also pure,
\begin{equation}
\rho(t) = | \psi(t) \rangle \langle  \psi(t) |
\; , \;\;\;
| \psi(t) \rangle = U_t | \psi_0 \rangle \; ,
\label{puree}
\end{equation}
cf.\ \era{rhot}{U}. It follows that the von Neumann entropy of the whole system is always zero,
\begin{equation}
S(t) = - \rho(t) \ln \rho(t) = 0 \; .
\label{S}
\end{equation}

Consider now the reduced state in partition $A$ obtained by tracing out partition $B$,
\begin{equation}
\rho_A(t) = {\rm Tr}_B \rho(t) \; .
\label{rhoA}
\end{equation}
Even if $\rho(t)$ is pure, in general $\rho_A(t)$ will be mixed. Furthermore, the entropy of the reduced state,
\begin{equation}
S_A(t) = - \rho_A(t) \ln \rho_A(t) \; ,
\label{SA}
\end{equation}
quantifies the entanglement (i.e., quantum correlations) between partitions $A$ and $B$, and is called the (bipartite) {\em entanglement entropy}. Note: (i) we obtain the same amount of entanglement if we trace out $A$ instead, i.e., $S_A = S_B$; and (ii) as long as $\rho(t)$ is pure, $S_A$ is a {\em measure} of entanglement, i.e., it tells not only if there is bipartite entanglement, but also quantifies it, and therefore can be used to quantify its change under a certain process, e.g., how entanglement grows in time under coherent evolution (see below).

We established above that $\rho(t)$ for the whole system $A+B$ equilibrates.  Thermalisation refers to the form that the reduced state $\rho_A(t)$ acquires at long times, from the point of view of expectation values of local observables in $A$,
\begin{equation}
\langle O_A(t) \rangle = {\rm Tr}_A \left[ O_A \rho_A(t) \right] \; .
\label{OA}
\end{equation}
Specifically,
\begin{equation}
\lim_{t \to \infty}
\langle O_A(t) \rangle = {\rm Tr}_A \left( O_A {\rm Tr}_B \rho_{\rm th} \right)
\; ,
\label{OA2}
\end{equation}
where $\rho_{\rm th}$ is a thermal density matrix for the whole system,
\begin{equation}
\rho_{\rm th} = e^{- \beta H } \; .
\label{rhoth}
\end{equation}
The inverse temperature $\beta$ is set by the energy of the initial state, which of course conserved under the evolution, that is, 
\begin{equation}
\beta \; {\rm s.t.}: \langle \psi_0 | H | \psi_0 \rangle =
{\rm Tr} \left( H \rho_{\rm th} \right) 
\; . 
\label{beta}
\end{equation}
This means that the only memory of the initial conditions that remains in the effective long time state of partition $A$ is that of the initial energy. All other initial details are forgotten. Thermalisation is thus the general setup for {\em quantum ergodicity} where the system can act as its own thermal reservoir. 
\changes{
For references see e.g., \cite{Calabrese2005,Barthel2008,Cramer2008,Fagotti2013,Essler2016,Alba2017}.
}

Quantum thermalisation can be seen as a consequence of the spectrum of a many-body system obeying the eigenstate thermalisation hypothesis (ETH)  \cite{Deutsch1991,Srednicki1994,Tasaki1998,Rigol2008}. When ETH holds, the spectrum of such a system is similar - in the large size limit - to that of a random matrix with the same symmetries of the Hamiltonian. In particular, matrix elements of local observables in the energy eigenbasis have the \changes{following form}, 
\begin{equation}
\langle E_n | \mathcal{O} | E_m \rangle = 
\delta_{nm} \, F(\bar{E}) 
+ 
e^{-S(\bar{E})/2} \, g(\bar{E},\omega) \, R_{nm}
\; ,
\label{RMT}
\end{equation}
where $\bar{E} = (E_n+E_m)/2$, $\omega = E_n-E_m$, $F(E)$ is some smooth function of $E$, $S(E)$ is the entropy of states with energy $E$, and $R_{nm}$ are Gaussian random numbers with zero mean and unit variance. It follows from \er{RMT} that 
for large size local observables become diagonal in the energy eigenbasis with random off-diagonal corrections that vanish exponentially with size. This means that expectation values at long times are well approximated by micro-canonical averages in the energy shell set by the initial conditions (or equivalently, canonical averages at the corresponding temperature) and thermalisation ensues.  Just like for  Gaussian random matrices, the spectrum of ETH obeying systems displays level repulsion, and the entanglement entropy of eigenstates in the bulk of the spectrum obeys a {\em volume law} (i.e., bipartite entanglement scales with the size of the smaller partition).  This volume law makes $S_A$, cf.\ \er{SA}, generically extensive in size, thus giving the entanglement entropy its thermal character.

\begin{figure*}[t]
\begin{center}
	\includegraphics[width=1.9\columnwidth]{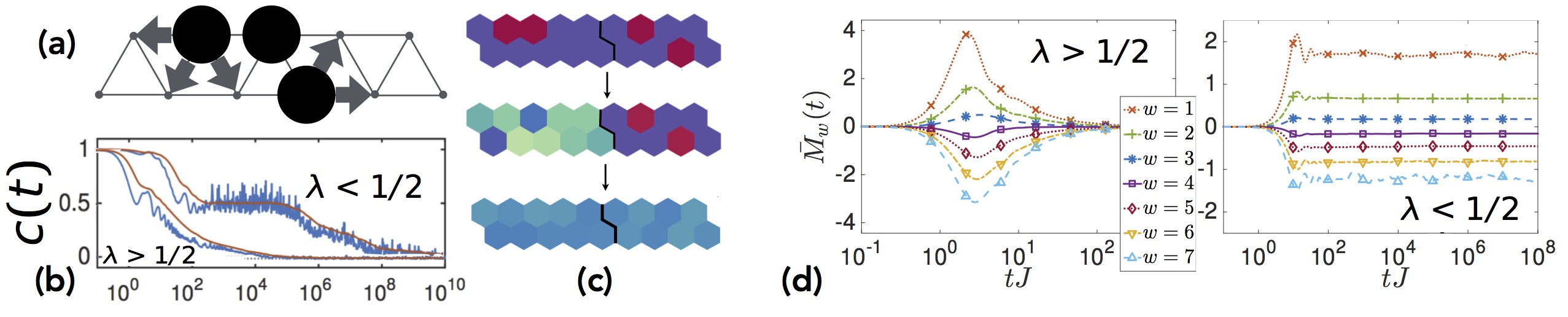}
\caption{
{\bf Slow quantum relaxation due to constraints:}
(a) Quantum constrained lattice gas, from Ref.\ \cite{Lan2017}.
(b) Change from fast to slow thermalisation and metastability in the relaxation of density correlations. 
(c) Slow thermalisation is associated with spatially heterogeneous growth of entanglement.
(d,e) Thermalisation vs.\ quasi-MBL in the quantum East model, from Ref.\ \cite{Horssen2015}. 
}
\label{fig7}
\end{center}
\end{figure*}

\subsection{Many-body localisation}

An exception to the above scenario are integrable systems \cite{Essler2016} which equilibrate to a generalised Gibbs ensemble \cite{Rigol2007,Vidmar2016}. A second notable exception, and the one to be discussed here, is that of (non-integrable) many-body quantum systems with quenched disorder that display many-body localisation (MBL) 
~\cite{Altshuler1997,Basko2006,Gornyi2005,Oganesyan2007,Znidaric2008,Pal2010,Bardarson2012,Serbyn2013,Huse2014,Andraschko2014,Yao2014,Serbyn2014,Laumann2014,De-Roeck2014,Scardicchio2015,Vasseur2015,Agarwal2015,Bar-Lev2015,Imbrie2016,Schreiber2015,Bordia2016,Smith2016}; for reviews see \cite{Nandkishore2015,Altman2015,Abanin2017}. 

A typical system that can undergo a transition from a thermal to a MBL phase is the following \cite{Pal2010}, cf.\ Fig.\ 6(b), 
\begin{equation}
H = -J 
\sum_{i=1}^{N} 
\left( c^\dagger_i c_{i+1} + c^\dagger_{i+1} c_{i} \right) 
+ V \sum_{i=1}^N n_i n_{i+1}
+ \sum_{i=1}^N h_i n_i 
 \; ,
\label{HMBL}
\end{equation}
for a system of spinless fermions in a one-dimensional lattice of size $N$. The first sum in \er{HMBL} corresponds to nearest neighbour hopping with rate $J$, and the second, to density-density interactions with coupling constant $V$. The third set of terms corresponds to a random local energies of strength that are random i.i.d.\ $h_i \in [-h, h]$. For $h=0$ the clean interacting problem thermalises for all values of $V>0$. In the non-interacting case, $V=0$, this is a typical system that displays Anderson localisation (AL) for all $h > 0$. 

In the clean case ($h=0$) dynamics evolved under \er{HMBL} thermalises.  In the presence of weak disorder, thermalisation is also achieved.  A signature of ergodicity is that time correlations decay (appropriately normalised) to zero. A typical test, which is also applicable in experiments, is to start from an initial product state with a well defined density profile, such as the following half-filling state
\begin{equation}
| \psi_0 \rangle = | \bullet \circ \bullet \circ \bullet \circ \bullet \circ \rangle \; ,
\label{CDW}
\end{equation}
sometimes also called a charge density wave. A simple observable which quantifies the extent to which the system remembers the initial patter is the difference in density between even and odd sites, or ``imbalance'' \cite{Schreiber2015}, 
\begin{equation}
\mathcal{I}(t) = \frac{N_o(t)-N_e(t)}{N_o(t)+N_e(t)} \; .
\label{Imb}
\end{equation}
Thermalisation implies forgetting details about initial conditions, and therefore the imbalance should go to zero at long times under ergodic conditions. (While the random fields in principle break translation invariance for any given sample, both for large systems and/or when averaging over disorder, this effect becomes negligible.) Figure 6(c) shows results from simulations of the system of \er{HMBL}: for small $h$ (red curve) 
the imbalance both becomes stationary and goes to zero within simulation times, compatible with thermalisation.  In contrast, when the random field exceeds a critical value, $h > h_c$, there is a transition into the MBL phase. In Fig.\ 6(c) this is manifested as the imbalance, while stationary, remains non-zero at long times; this shows that for strong disorder the system maintains memory of the initial state, an indication of non-ergodicity.  For reference, Fig.\ 6(c) also shows the non-interacting case of AL. 

The MBL state has been well characterised by now, both in terms of spectral properties and dynamics \cite{Nandkishore2015,Altman2015,Abanin2017}. MBL implies a breakdown of ETH. Under MBL conditions the eigenstates in the bulk of the spectrum have low entanglement - similar to low lying states of gapped Hamiltonians (i.e., displaying area rather than volume law entanglement), and there is no level repulsion.  In fact, \changes{an} MBL system has an extensive number of emergent local conservation laws - something which is believed to be true in general and has indeed been proven rigorously for certain specific problems. On the dynamical side, a key distinction between an interacting system in the thermal and MBL phase is the growth of entanglement.  This is illustrated in Fig.\ 6(d) for the model of \er{HMBL} under the same conditions as Fig.\ 6(c): the initial product state \er{CDW} has zero bipartite entanglement; in the thermal phase, $h < h_c$, the half-chain entanglement entropy grows rapidly (red curve) towards a value which is extensive in system size; the growth on the MBL side, $h > h_c$, in contrast is much slower, first achieving a sub-extensive level similar to that of AL, and subsequently progressing logarithmically in time; while the asymptotic value is extensive it is much lower than that of the thermal phase, as a consequence of MBL eigenstates obeying area law, which means that the extensivity is a consequence of the superposition that is reached at long times.

\subsection{Slow quantum relaxation due to constraints}

A key distinction between MBL and classical glasses is that the former are disordered systems while in the latter disorder is absent.  Whether MBL can exist in translationally invariant systems is still debated \cite{Carleo2012,Horssen2015,Schiulaz2015,Papic2015,Barbiero2015,Yao2016,Prem2017,Smith2017,Yarloo2017,Mondaini2017}.  Even if it were the case that strict asymptotic quantum non-ergodicity could not exist in the absence of disorder, an interesting question to address is whether the standard mechanisms for glassy slowing down - in particular kinetic constraints encoding steric restrictions to local motion - can give rise to analogous slow relaxation in closed quantum systems. Figure 7 summarises some of the findings from two recent papers that address this precise issue by considering two quantum models inspired by classical KCMs. 

The first model \cite{Lan2017} is a quantum version of a {\em constrained lattice gas} \cite{Ritort2003}.  
It consists of hard-core particles moving on a 1D strip of a triangular lattice of $L$ sites and occupation $N$; see Fig.\ 7(a). The Hamiltonian - described in terms of spin-$1/2$ operators - is
\begin{align}
	H_{\rm QLG} = & -\tfrac{1}{2} \sum_{\langle i, j\rangle} 
	\hat{C}_{ij} \left\{
	\lambda
	\left( 
	\sigma^{+}_{i} \sigma^{-}_{j} + \sigma^{+}_{j} \sigma^{-}_{i}
	\right) 
	\right.
\nonumber
\\
	& 
	\left.
	- (1-\lambda) 
	\left[
	n_{i} (1 - n_{j}) + n_{j}(1 - n_{i})
	\right] 
	\right\}
	\; .
	\label{Htlg}
\end{align}
The operator $\hat{C}_{ij} = 1 - \prod_k n_k$ plays the role of a kinetic constraint: the product is over all sites $k$ that are common neighbours of both sites $i$ and $j$, and as Fig.\ 7(a) shows, it allows motion only when at least one of the common neighbouring sites is empty, thus mimicking excluded volume restrictions. The model conserves density but has no particle-hole symmetry.  The effect of the constraints is only significant for large fillings, where many moves that would be possible in the unconstrained problem are blocked.

Consider unitary evolution under \er{Htlg}.  We take as initial states $\ket{\psi_0}$ product states corresponding to classical configurations, discarding those with only isolated vacancies (which are disconnected under $H$). To quantify relaxation consider the two-time (connected and scaled) density autocorrelator
\begin{equation}
c(t) = \frac{1}{L} \sum_i 
\frac{\langle \psi_0 | n_i(t) n_i(0) |\psi_0\rangle}
{\phi(1-\phi)}
-
\frac{\phi}{(1-\phi)} 
\label{ct}
\; ,
\end{equation}
where \(n_i(t)\) is the Heisenberg-picture number operator and $\phi=N/L$ is the filling fraction.  Since $|\psi_0 \rangle$ is a product state, $\langle \psi_0 | n_i(t) n_i(0) |\psi_0\rangle$ reduces to the expectation value $\langle n_i(t) \rangle$ for initially occupied sites $i$, so that $c(t)$ plays a role similar to the imbalance, cf.\ \er{Imb}, in that it quantifies memory of the initial density profile. Figure 7(b) shows $c(t)$ (and a running time average to smooth out short time fluctuations).  The key observation is that for $\lambda > 1/2$, such that the kinetic terms in $H$ dominates over the potential terms, thermalisation is fast.  In sharp contrast, for $\lambda < 1/2$, where potential energy dominates over the kinetic energy, there is a pronounced separation of timescales, with fast decay to a nonzero plateau, and eventual thermalisation only at much longer times. Compare this two-step relaxation to time-correlators in classical glassy systems, Fig.\ 1(b).  This shows that, just like in the classical case, kinetic constraints can also induce slow and cooperative relaxation in quantum systems under the right conditions.  Furthermore, for $\lambda < 1/2$, metastability and slow thermalisation is accompanied by spatially heterogenous growth of entanglement, as shown in Fig.\ 7(c).  Again this is reminiscent of dynamic heterogeneity in the classical case, cf.\ Fig.\ 1(d).

The second model is the {\em quantum East model} \cite{Horssen2015}, defined in terms of quantum spins in a one-dimensional lattice with Hamiltonian, 
\begin{equation}
H_{\rm East} = - \sum_i n_{i-1} \left[ \lambda \sigma_i^x - \left( 1 - \lambda \right) \right] \; .
\label{qEast}
\end{equation}
This Hamiltonian is (up to a sign) the same operator as the tilted generator of the classical East model, \er{WEasts}, at temperature $T=\infty \Rightarrow \epsilon=1$.  As in the classical case, the constraint only allows spin flips on sites whose left neighbour has a projection on the spin-up state.  As in the previous model, the parameter $\lambda$ biases the weight of the kinetic versus the potential energy terms in $H_{\rm East}$.  Figure 7(c) illustrates the basic dynamics of this model. For $\lambda > 1/2$ the system thermalises: the left panel shows the total longitudinal magnetisation starting from all possible product states at half filling (note that in this model filling is not conserved); these states all have the same expectation value of the energy, so from ETH we expect that asymptotically the observable reaches the same value, as they do.  In contrast, for $\lambda < 1/2$ 
dynamics is orders of magnitude slower and relaxation is not achieved on numerical accessible timescales: the observable retains memory of details of the initial conditions for all accessible times, see right panel.  While this does not prove asymptotic non-ergodicity, and might indeed be quasi-MBL (MBL like dynamics for long times, with eventual thermalisation beyond times that can be reached numerically), it again proves that kinetic constraints can lead to dynamical arrest over very long times in analogy with what occurs in glasses.

\begin{figure}[t]
	\includegraphics[width=\columnwidth]{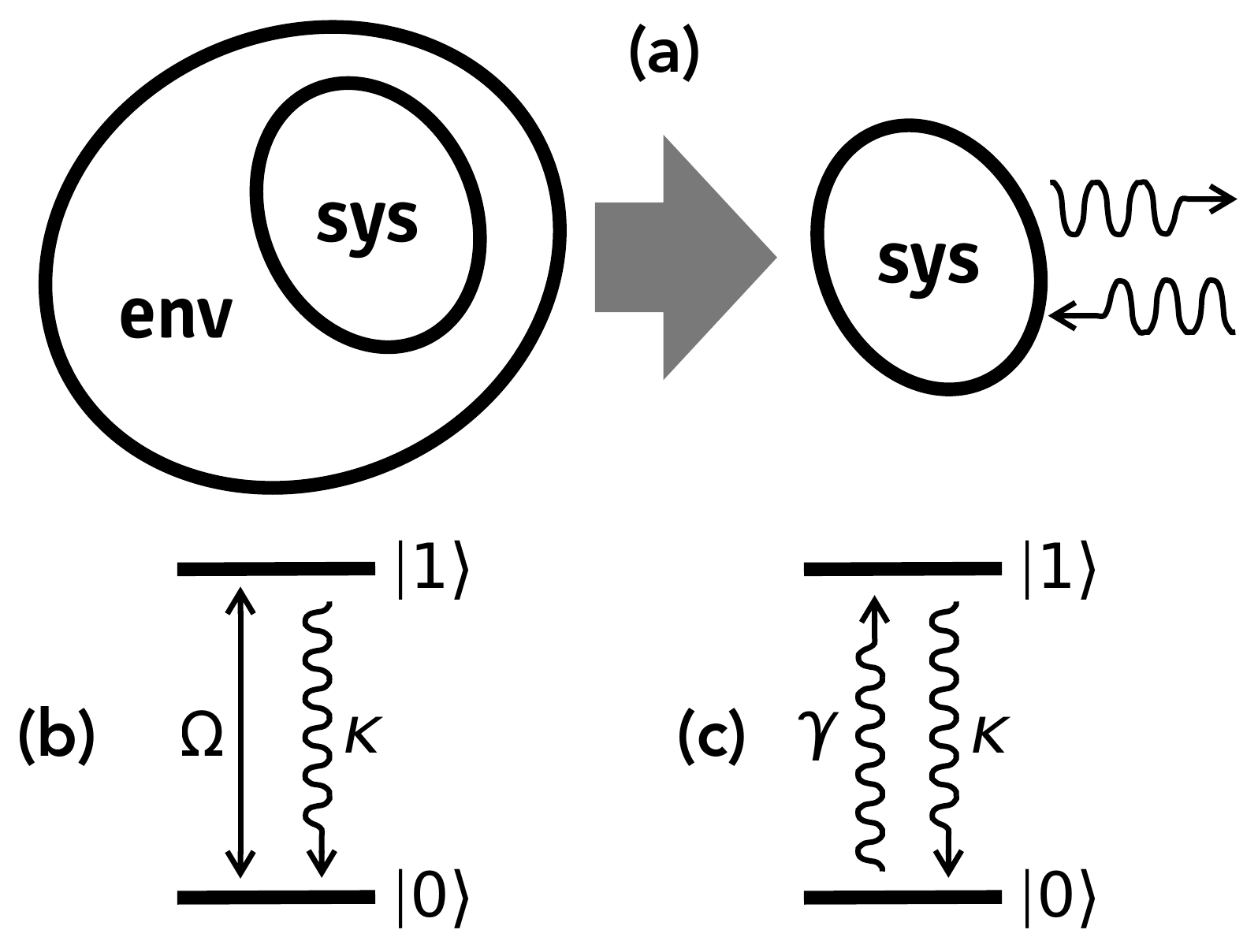}
\caption{
{\bf Open quantum systems:}
(a) Reduction of the full system+environment unitary dynamics to an effective quantum Markovian description of the system. 
(b) Coherently driven 2-level system at zero temperature.
(c) Classical 2-level system at non-zero temperature.
}
\label{fig8}
\end{figure}

\section{Non-equilibrium in open quantum systems}
\label{Lec4}

\subsection{Markovian open quantum dynamics} 

Consider a closed quantum system, Fig.\ 8(a), composed of the subsystem of interest (denoted {\bf sys} in the figure), and rest that we will call the {\em environment} (denoted {\bf env} in the figure). The total system evolves unitarily according to a total Hamiltonian, 
\begin{equation}
H_{\rm tot} = H + H_{\rm env} + H_{\rm int} \; ,
\label{Htot}
\end{equation}
where $H$ acts on the system, $H_{\rm tot}$ acts on the environment, and $H_{\rm int}$.  A general form for the interaction Hamiltonian is
\begin{equation}
H_{\rm int} = \sum_{\mu=1}^{N_{\rm J}} \left( J_\mu b_\mu^\dagger + J_\mu^\dagger b_\mu \right) \; ,
\label{Hint}
\end{equation}
where $J_\mu,J_\mu^\dagger$ are operators that act on the system (which we call {\em jump operators} below), and $b_\mu,b_\mu^\dagger$ are operators on the environment.  For the environment we assume the diagonal form, 
\begin{equation}
H_{\rm env} = \sum_{\mu=1}^{N_{\rm J}} \omega_\mu b_\mu^\dagger b_\mu 
\; .
\label{Henv}
\end{equation}
We will further assume that the environment is large, that correlation times in the environment are short compared to relevant timescales of the system, and that the system-environment coupling is weak.  Under these conditions it is possible to trace out the environment and obtain an effective Markovian description of the dynamics of the system in terms of a {\em quantum master equation} (QME) \cite{Lindblad1976,Gorini1976,Plenio1998,Breuer2002,Gardiner2004b,Daley2014}, 
\begin{equation}
\partial_t \rho = -i [H , \rho] + \sum_\mu J_\mu \rho J_\mu^\dagger
- \frac{1}{2} \{ J_\mu^\dagger J_\mu , \rho \} 
\; .
\label{QME}
\end{equation}
On the r.h.s.\ of \er{QME} the first term describes the part of the dynamics due to the system Hamiltonian - i.e., the coherent part of the evolutions - while the rest corresponds to the effect of the environment on the system under the Markovian approximation - i.e., the dissipative part of the evolution, cf.\ Fig.\ 8(a).  We can write the QME as 
\begin{equation}
\partial_t \rho = \L(\rho)
\; ,
\label{QME2}
\end{equation}
where the generator $\L$ is a {\em superoperator} (an operator acting on matrices) \cite{Breuer2002,Gardiner2004b}, 
\begin{equation}
\L(\cdot)  = -i [H , (\cdot)] + \sum_\mu J_\mu (\cdot) J_\mu^\dagger
- \frac{1}{2} \{ J_\mu^\dagger J_\mu , (\cdot) \} 
\; .
\label{L}
\end{equation}
The QME equation is often called the Lindblad equation, and the generator $\L$ the Lindbladian.  Formally integrating \er{QME2} we obtain the state of the system at any time $t > 0$ given the initial state $\rho_0$,
\begin{equation}
\rho(t) = e^{t \L} (\rho_0) \; .
\label{rhot}
\end{equation}
In contrast to unitary evolution, the dynamics generated by $\L$ is dissipative, and as such generically the state of the system at long times tends to a stationary state,
\begin{equation}
\lim_{t \to \infty} \rho(t) = \rho_{\rm ss} \; ,
\label{rhoss}
\end{equation}
which in general is unique. (Just like in the classical case, under certain conditions, dynamics can be reducible with the existence of more than one stationary state \cite{Albert2016}.)

The form \er{QME} is the general form for a local equation that gives rise to evolution which preserves the properties of the density matrix - \er{rhot} is a {\em completely positive trace preserving} (CPTP) map. In fact, it is easy to show that if $\rho_0$ is a valid density matrix, then under \er{rhot},
\begin{equation}
\rho(t)^\dagger = \rho(t) 
\; , \;\;\;
{\rm Tr}[\rho(t)] = 1
\; , \;\;\;
{\rm Tr}[\rho(t)^2] \leq 1 
\;\;
\forall t
\; ,
\label{props}
\end{equation}
where we have assumed ${\rm Tr}[\rho_0] = 1$. 

The structure and properties of the QME \ers{QME}{L} are very similar to that of the classical ME \era{ME1}{ME2}.  Table 2 provides a dictionary between the classical ME and the QME.  The QME \er{QME} is a deterministic equation for the density matrix, a quantum dissipative analog of the deterministic equation \er{ME1} for the probability in the classical case.  Both the classical Markov generator $\W$  and the Lindbladian $\L$ have zero as their largest eigenvalue.  The corresponding right eigenvector in the classical case is the stationary state probability $| {\rm ss} \rangle$, while in the quantum case the right eigenmatrix is the stationary state $\rho_{\rm ss}$. The corresponding left eigenvector/eigenmatrix are the flat state $\fs$ for $\W$ and the identity matrix $\I$ for $\L$. These left eigenstates are always the same irrespective of the form of $\W$ and $\L$, as this is the statement of probability conservation.  In terms of $\L$, ``action to the left'' is defined in terms of the adjoint $\L^\dagger$, where
\begin{equation}
\L^\dagger(\rho) = i [H , \rho] + \sum_\mu J_\mu^\dagger \rho {J_\mu}
- \frac{1}{2} \{ J_\mu^\dagger J_\mu , \rho \} 
\; ,
\label{Ldag}
\end{equation}
and it is easy to show that $\L^\dagger(\I)=0$ holds. The structure of the operator $\W$ and superoperator $\L$ is also analogous.  The classical generator has off diagonal terms in the configuration basis which are positive and encode the possible jumps between configurations and their probabilities.  Each of these operators is of the form, $W({C \to C'}) |C' \rangle \langle C|$.  The corresponding operators in the quantum case are the jump operators $J_\mu$, acting on the density matrix as $J_\mu (\cdot) J_\mu^\dagger$.  The diagonal part of the classical generator $\W$ is the escape rate operator, $\sum_{C} R({C}) |C \rangle \langle C|$, with each entry corresponding to the escape rate of each configuration. A similar role is played in the quantum case by $-i H_{\rm eff}$, where the {\em effective Hamiltonian} is defined as
\begin{equation}
H_{\rm eff} = H - \frac{i}{2} \sum_\mu J_\mu^\dagger J_\mu 
\; . \label{Heff}
\end{equation}
In fact, if we rewrite the Lindbladian \er{L} as,
\begin{equation}
\L(\cdot)  = \sum_\mu J_\mu (\cdot) J_\mu^\dagger - i
\left[
H_{\rm eff} (\cdot) - (\cdot) H_{\rm eff}^\dagger
\right]
\; ,
\label{L2}
\end{equation}
we see that $\L$ has a structure analogous to that of \er{W}.

An interesting observation is that the classical ME is contained in the QME: for example, if $H=0$, and all the jump operators are rank-1, i.e., of the form $J_\mu \propto |C' \rangle \langle C|$, the dynamics of the diagonal of the density matrix decouples from the dynamics of the off-diagonal elements, and the diagonal dynamics is equivalent to that of a classical probability vector; this point is illustrated below.

\changes{
For a small sample of applications of this formalism to problems in open quantum dynamics see e.g.\ \cite{Molmer1993,Plenio1998,Daley2014,Prosen2008,Prosen2011,Prosen2011b,Ates2012,Karevski2013,Lesanovsky2013,Cai2013,Znidaric2014,Macieszczak2016}.
}

\begin{table*}[t]
\begin{center}
\begin{tabular}{rcccc}
\hline \hline \\
 && {\bf classical} && {\bf open quantum } \\
\\ \hline \\
 master equation: && 
$\partial_{t} |P(t) \rangle = \W |P(t) \rangle$  && 
$\partial_t \rho = \L(\rho)$ \\
\\ \hline \\
 right zero-eigenstate: && 
 $\W | {\rm ss} \rangle = 0$ && 
 $\L(\rho_{\rm ss})=0$ \\
\\ \hline \\
 left zero-eigenstate: && 
 $\fs \W = 0$ && 
 $\L^\dagger(\I) = 0$  \\
\\ \hline \\
jump operators: && 
 $W({C \to C'}) |C' \rangle \langle C|$ && 
 $J_\mu$  \\
\\ \hline \\
  escape rate operator: && 
 $\sum_{C} R({C}) |C \rangle \langle C|$ && 
 $i H_{\rm eff} = i H + \frac{1}{2} \sum_\mu J_\mu^\dagger J_\mu$  \\
 \\ 
 \hline
 \hline
\end{tabular}
\caption{Comparison between classical and quantum continuous time Markovian dynamics.}
\end{center}
\end{table*}

\subsubsection{Example: 2-level system at $T=0$}

Consider the problem illustrated in Fig.\ 8(b). It depicts a system with two levels $|0\rangle$ and $|1 \rangle$ (i.e., a qubit) under the action of a driving Hamiltonian
\begin{equation}
H = \Omega \sigma_x \; ,
\label{H2}
\end{equation}
where $\sigma_x = |0\rangle \langle 1| + |1\rangle \langle 0|$.  (This could model, e.g., an atom where two atomic levels are driven by a laser on resonance with the line between them, described in the rotating wave approximation with $\Omega$ the strength of the drive, or Rabi frequency.) The system of Fig.\ 8(b) is also subject to a dissipative process described by a single jump operator
\begin{equation}
J = \sqrt{\kappa} \sigma_- 
\; ,
\label{J2}
\end{equation}
where $\sigma_- = |0\rangle \langle 1|$ (and $\sigma_+ = \sigma_-^\dagger$).  This models for example the interaction with a zero temperature environment where the system can only emit, say photons, into the environment, but not absorb from it. The rate for these emissions is $\kappa$. 
The Lindbladian for this 2-level problem explicitly reads,
\begin{equation}
\L (\cdot) =  - i \Omega [ \sigma_x , (\cdot)] + \kappa \sigma_- (\cdot) \sigma_+
- \frac{\kappa}{2} \{ n , (\cdot) \} 
\; ,
\label{L2}
\end{equation}
where $n = \sigma_+ \sigma_-$. Given that $\rho$ is a $2\times2$ matrix, the superoperator $\L$ can be thought of as a $4\times4$ matrix and diagonalised explicitly.  For simplicity let us consider the case where $\kappa = 4 \Omega$, for which expressions are more compact (as certain square roots in the general solution vanish).  The four eigenvalues of $\L$ are,
\begin{equation}
\lambda_0 = 0 
\; , \;\;
\lambda_1 = -2 \Omega 
\; , \;\;
\lambda_2 = \lambda_3^* = -3 \Omega - i \sqrt{3} \Omega  
\; , \;\;
\label{evals2}
\end{equation}
where we have labeled the eigenvalues by decreasing real part. The first eigenvalue vanishes, as expected, and corresponds to the stationary state. All other eigenvalues have negative real parts, indicating that their modes decay exponentially in time.  For each eigenvalue there are right and left eigenmatrices, i.e., 
\begin{equation}
\L(R_n) = \lambda_n R_n 
\; , \;\;
\L^\dagger(L_n) = \lambda_n L_n 
\; .
\label{LR}
\end{equation}
For $\L$ of \er{L2} the right eigenmatrices are
\begin{align}
&
R_0 = 
\left(
\begin{array}{rr} 
\frac{1}{6} & - \frac{i}{3} \\ \frac{i}{3} & \frac{5}{6} \\
\end{array}
\right) 
\; , \;
R_1 = 
\left(
\begin{array}{rr} 
0 & \frac{1}{2} \\ 0 & \frac{1}{2} \\
\end{array}
\right)
\; , \; \nonumber \\
&
R_2 = R_3^\dagger =
\left(
\begin{array}{cc} 
- \frac{1 + i \sqrt{3}}{12} & \frac{i}{6} \\ 
- \frac{i}{6} & \frac{1 + i \sqrt{3}}{12} \\
\end{array}
\right) 
\; ,
\label{rr2}
\end{align}
and the left eigenmatrices are
\begin{align}
&
L_0 = \I 
\; , \;
L_1 = \sigma_x
\; , \; \nonumber \\
&
L_2 = L_3^\dagger =
\left(
\begin{array}{cc} 
1+ i 2\sqrt{3} & \frac{i 2 \sqrt{3}}{\sqrt{3} - i} \\ 
- \frac{i 2 \sqrt{3}}{\sqrt{3} - i} & 1 \\
\end{array}
\right) 
\; ,
\label{ll2}
\end{align}
and we have chosen a normalisation such that
\begin{equation}
\Tr \left( L_n \cdot R_m \right) = \delta_{nm} \, .
\label{LRnorm}
\end{equation}
We can use the above for a spectral decomposition of the generator so that 
\er{rhot} becomes
\begin{equation}
\rho(t) = \sum_n e^{t \lambda_n} \Tr \left( L_n \cdot \rho_0 \right) R_n 
\; . 
\label{rhotLR}
\end{equation}
From here we see that the stationary state coincides with $R_0$, and in the particular case of the 2-level system of Fig.\ 2(b) we have
\begin{equation}
\rho_{\rm ss} = R_0 = \left(
\begin{array}{rr} 
\frac{1}{6} & - \frac{i}{3} \\ \frac{i}{3} & \frac{5}{6} \\
\end{array}
\right) 
\; .
\label{rhoss2}
\end{equation}
\Er{rhoss2} says that in the stationary state the average occupation of the upper state is $\langle n \rangle = 1/6$ and that of the lower state $1 - \langle n \rangle = 5/6$, and there are also steady state coherences so that $\langle \sigma_y \rangle = 2/3 \neq 0$.

\subsubsection{Example: classical 2-level system}

It was mentioned above that a classical master equation can be embedded in the QME.  Consider the example of the same 2-level system, but now we replace the coherent term in the dynamics by a dissipative jump from the the $|0\rangle$ to the $|1\rangle$ state, see Fig.\ 8(c).  Of course this corresponds to the dynamics of classical Ising spin, but let us consider how such system in described in the QME framework.  In this case we have the following Hamiltonian and jump operators,
\begin{equation}
H = 0 
\; , \;
J_1 = \sqrt{\kappa} \sigma_- 
\; , \;
J_2 = \sqrt{\gamma} \sigma_+ 
\; ,
\label{J2c}
\end{equation}
leading to the Lindbladian,
\begin{equation}
\L (\cdot) =  
\kappa \sigma_- (\cdot) \sigma_+
+
\gamma \sigma_+ (\cdot) \sigma_-
- \frac{\kappa-\gamma}{2} \{ n , (\cdot) \} 
- \gamma (\cdot)
\; .
\label{L2c}
\end{equation}
If we decompose the density matrix as 
\begin{equation}
\rho =
\left(
\begin{array}{cc} 
p_1 & x - i y \\ 
x + i y & p_0 \\
\end{array}
\right) 
\; , \;
\label{rho2c} 
\end{equation}
where $p_{0,1}$ are the probabilities of occupation of the states, with $p_0 + p_1 = 1$, and $x,y$ the coherences, we find that the QME, \er{QME2}, splits into two disconnected parts,
\begin{equation}
\partial_t \rho = \L(\rho) 
\Rightarrow
\left\{
\begin{array}{l}
\partial_t \left(
\begin{array}{c} 
p_1 \\ 
p_0 \\
\end{array}
\right) 
=
\left(
\begin{array}{c} 
\gamma p_0 - \kappa p_1 \\ 
\kappa p_1 - \gamma p_0 \\
\end{array}
\right) 
\\
\\
\partial_t \left(
\begin{array}{c} 
x \\ 
y \\
\end{array}
\right) 
=
-\frac{\kappa+\gamma}{2}
\left(
\begin{array}{c} 
x \\ 
y \\
\end{array}
\right) 
\\
\end{array}
\right.
\; ,
\label{QME2c}
\end{equation}
and the probabilities and coherences decouple.  If the initial state $\rho_0$ is diagonal, then the dynamics described by the Lindbladian \er{L2c} is the same as that of a classical generator 
\begin{equation}
\W = \left(
\begin{array}{cc} 
- \kappa & \gamma \\ 
\kappa & - \gamma \\
\end{array}
\right) 
\; , 
\label{W2c}
\end{equation}
acting on a probability vector $(p_1, p_2)$.  If the initial state has any coherences, from \er{QME2c} it follows that they decay exponentially fast, taking the system into the classical subspace.

\begin{figure}[t]
	\includegraphics[width=\columnwidth]{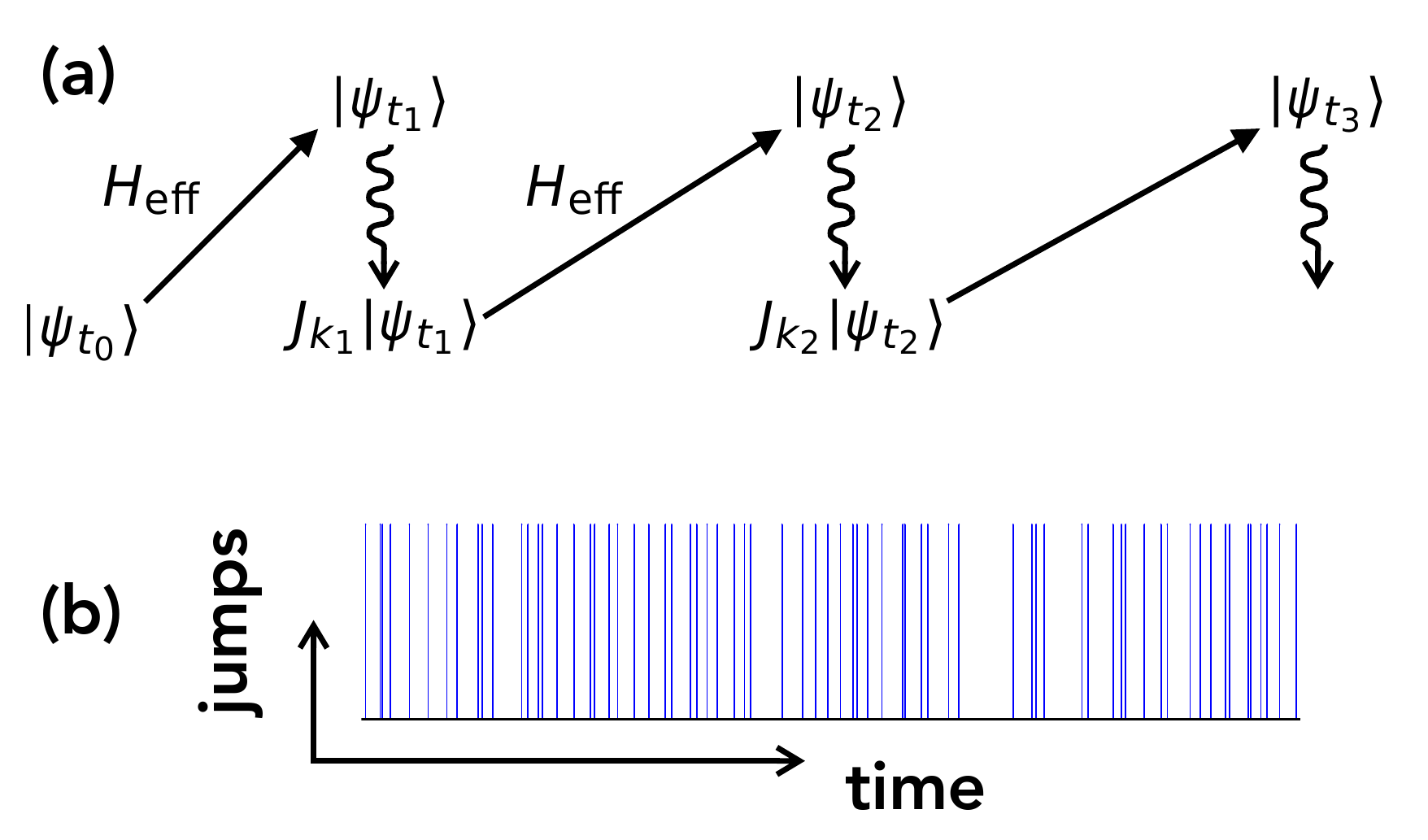}
\caption{
(a) Sketch of a quantum trajectory from a quantum jump unravelling of the QME.
(b) Sketch of a quantum jump trajectory.}
\label{fig9}
\end{figure}

\subsection{Quantum jump trajectories} 

The Lindbladian, \er{L}, generates a continuous {\em quantum Markov chain}  \cite{Breuer2002,Gardiner2004b}.  Just like the classical ME is associated to stochastic trajectories between configurations, cf.\ Fig.\ 4(b), the QME can be {\em unravelled} in terms of stochastic quantum trajectories \cite{Belavkin1999,Dalibard1992,Plenio1998,Breuer2002,Gardiner2004b}.  As in the classical case, averaging the state of the system at any one time over the set of stochastic trajectories recovers the deterministic evolution of the density matrix under the QME \er{QME}. 

Quantum unravellings are not unique.  They depend on the choice of observational basis in the environment that is chosen: for example for situations common in quantum optics, different unravellings could be those given by counting photons as opposed to homodyne measurement of photocurrents
(associated with quantum trajectories described in terms of quantum stochastic differential equations, where the quantum noise due to the environment enters as a quantum Wiener process). The particular class of unravellings we will consider, which makes the analogy to the classical case more direct, is that based on {\em quantum jumps}. 

Figure 9(a) sketches a quantum trajectory associated with the QME \er{QME}, corresponding to a quantum jump unravelling. Let us assume for simplicity that we start from a pure state $|\psi_{t_0} \rangle$ at time $t_0$. A trajectory will correspond to the evolution of a pure state $| \psi_t \rangle$ given by periods of deterministic evolution, generated by $H_{\rm eff}$, cf.\ \er{Heff}, punctuated by sudden jumps of the wavefunction due to the action of the jump operators $J_\mu$ occurring at random times.  The periods between quantum jumps are analogous to the periods between classical jumps in the case of classical Markov chains, compare Fig.\ 9(a) with Fig.\ 4(b).  The evolution between jumps is given by
\begin{equation}
| \psi_{t_k + \Delta t} \rangle = e^{-i \Delta t H_{\rm eff}} | \psi_{t_k} \rangle
\; , 
\label{psieff}
\end{equation}
where $| \psi_{t_k} \rangle$ is the state {\em after} the previous quantum jump that occurred at a time we denote $t_k$. As mentioned above, the operator $H_{\rm eff}$ plays the role of the escape rate operator for the quantum trajectories.  In contrast to the classical case, $H_{\rm eff}$ is in general non-diagonal, and as such the state evolves in time between jumps.   

Since in general $H_{\rm eff} \neq H_{\rm eff}^\dagger$, the evolution in \er{psieff} is non-unitary, and as such the norm of the state is not preserved. 
This non-conservation of the norm is associated with the survival time before the next quantum jump occurs. In particular, if the last jump occurred at time $t_k$, the probability of surviving (i.e., having no jump) until time $t_k + \Delta t$ is related to the norm of \er{psieff} by,
\begin{align}
P_{| \psi_{t_k} \rangle}(\Delta t) & =
\frac{\norm{| \psi_{t_k + \Delta t} \rangle}^2}
{\norm{| \psi_{t_k} \rangle}^2}
\nonumber \\
& =
\frac{
\langle \psi_{t_k} | e^{i \Delta t H_{\rm eff}^\dagger} 
e^{-i \Delta t H_{\rm eff}} | \psi_{t_k} \rangle
}
{
\langle \psi_{t_k}  | \psi_{t_k} \rangle
}
\; .
\label{Pw}
\end{align}
This expression should be compared to the classical survival in a configuration which is exponential of the escape rate, $P(\Delta t | C) = e^{- \Delta t R(C)}$. 

After the no-jump evolution \er{psieff} from the time of the last jump at $t_k$ a new jump occurs at $t_{k+1} = t_k + \Delta t$.  The wavefunction changes by direct application of the jump operator to the state before the jump, i.e., 
\begin{equation}
| \psi_{t_{k+1}} \rangle_{\rm after} = J_{\mu_{k+1}} | \psi_{t_{k+1}} \rangle
\; . 
\label{psiJ}
\end{equation}
The change in the state occurs instantaneously, and thus the time index is the same before and after the jump, cf.\ Fig.\ 9(a).  The probability of it being jump $\mu$ out of the $N_{\rm J}$ possible jumps is given by 
\begin{equation}
p_{\mu}(| \psi_{t_{k+1}} \rangle ) =
\frac{
\langle \psi_{t_{k+1}} | J_\mu^\dagger J_\mu | \psi_{t_{k+1}} \rangle
}
{
\sum_{\nu=1}^{N_{\rm J}} \langle \psi_{t_{k+1}} | J_\nu^\dagger J_\nu | \psi_{t_{k+1}} \rangle
}
\; , 
\label{pmu}
\end{equation}
where $| \psi_{t_{k+1}} \rangle$ is the state just before the jump, \er{psiJ}.

Each jump is associated with an emission event observed in the environment, while the periods between jump correspond to periods where no emission is observed.  The time record of these observed emissions is called a {\em quantum jump trajectory}. 

Let us consider the 2-level system of Example 1 above (for the choice of parameters $\kappa = 4 \Omega$). Let us assume for simplicity that the initial state is the state $| 0 \rangle$.  From \era{H2}{J2} we get the effective Hamiltonian, cf.\ \er{Heff},
\begin{equation}
H_{\rm eff} = \Omega \sigma_x - \frac{\kappa}{2} n 
\; .
\label{Heff2}
\end{equation}
The survival probability between jump starting from state $| 0 \rangle$ is then
\begin{equation}
P_{| 0 \rangle}(\Delta t) =
e^{-2 \Omega \Delta t} 
\left[ 1 + 2 \Omega \Delta t + 2 (\Omega \Delta t)^2 \right]
\; .
\label{Pw2}
\end{equation}
Given that there is a single jump we have for \er{psiJ}
\begin{equation}
| \psi_{t_{k+1}} \rangle_{\rm after} = J | \psi_{t_{k+1}} \rangle
\propto | 0 \rangle
\; ,
\label{psiJ2}
\end{equation}
so that after each quantum jump the system is reverted to the initial state. In terms of the states after the jumps this corresponds to a renewal process,
\begin{equation}
| 0 \rangle \xrightarrow{\Delta t_1} 
| 0 \rangle \xrightarrow{\Delta t_2}
| 0 \rangle \cdots
\; , 
\label{ren2}
\end{equation}
which, in contrast to classical renewals, has waiting times $\Delta t$ that are not exponentially distributed, cf.\ \er{Pw2}. The associated quantum jump trajectories are a history of the times at which the single kind of quantum jump occurs, such as the sequence of clicks sketched in Fig.\ 9(b).

\subsection{Dynamical large deviations} 

We can generalise \cite{Garrahan2010} the LD approach of Lecture \ref{Lec2} to study ensembles of quantum jump trajectories (QJTs).  In particular, we wish to classify QJTs through dynamical observables such as the total number of jumps $K$ in a trajectory (e.g., the total number of photons emitted). \Ers{PKLD}{LT} generalise straightforwardly.  The SCGF can also be calculated from the largest eigenvalue of a tilted generator: if the observable of interest for a QJT $\omega_t$ is defined as 
\begin{equation}
K(\omega_t) = \sum_\mu \alpha_\mu K_\mu(\omega_t) 
\; , 
\label{Koq}
\end{equation}
where $K_\mu(\omega_t)$ is the total number of jumps of kind $\mu$ in trajectory $\omega_t$, then the associated tilted operator reads,
\begin{align}
\L_s(\cdot)  = -i [H , (\cdot)] + \sum_\mu e^{-s \alpha_\mu} J_\mu (\cdot) J_\mu^\dagger
& 
\nonumber \\
- \frac{1}{2} \{ J_\mu^\dagger J_\mu , (\cdot) \} &
\; .
\label{Ls}
\end{align}
Note that $K$ of \er{Koq} is a ``counting operator'' - cf.\ \er{Kgen} when $\beta_C = 0$ - and thus the tilting affects only the jump terms in $\L_s$.

\begin{figure}[t]
	\includegraphics[width=\columnwidth]{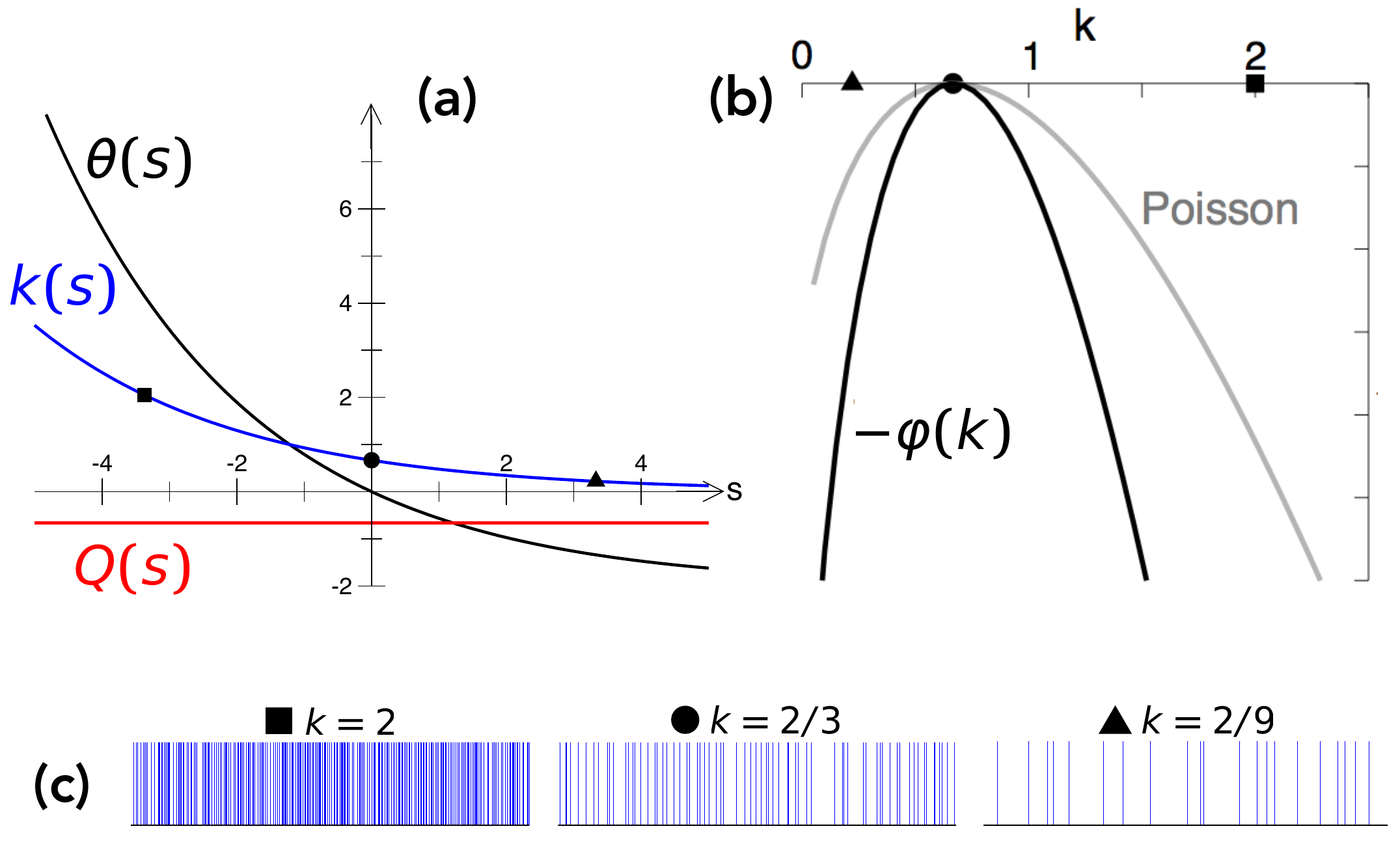}
\caption{
{\bf Large deviations in the 2-level system:}
(a) SCGF, activity and Q parameter as a function of $s$, for $\kappa = 4 \Omega$.  From Ref.\ \cite{Garrahan2010}.
(b) (Minus the) rate function of the activity (black). A Poisson rate function with same average is shown (grey) form comparison.
(c) Examples of trajectories at various values of $s$.  
}
\label{fig10}
\end{figure}

\subsubsection{Example: 2-level system} 

Let us go back to the example of 2-level system of section 4.1.1.  As an observable we consider the total number of jumps. The corresponding tilted operator reads \cite{Garrahan2010},
\begin{equation}
\L_s (\cdot) =  - i \Omega [ \sigma_x , (\cdot)] + e^{-s} \kappa \sigma_- (\cdot) \sigma_+
- \frac{\kappa}{2} \{ n , (\cdot) \} 
\; .
\label{L2s}
\end{equation}
Fixing again $\kappa = 4 \Omega$ for simplicity, we can obtain immediately the SCGF as the largest eigenvalue of \er{L2s}, 
\begin{equation}
\theta(s) = 2 \Omega \left( e^{-s/3} - 1 \right) 
\; .
\label{theta2}
\end{equation}
From a Legendre transform like \er{LT} we obtain the corresponding rate function
\begin{equation}
\varphi(k) = 3 \left[ k \ln \left( \frac{3 k}{2 \Omega} \right) -
\left( k - \frac{2 \Omega}{3} \right) \right] 
\; .
\label{phi2}
\end{equation}
The SCGF \er{theta2} and the rate function \er{phi2} are shown in Fig.\ 10(a,b). 

The moments of the number of jumps are obtained from the derivatives of $\theta(s)$ evaluated at zero.  For example, for the average number of jumps we get, 
\begin{equation}
\frac{\langle K \rangle}{t} = - \left. \frac{d \theta(s)}{ds} \right|_{s=0} = \frac{2}{3} \Omega
\; ,
\label{K2} 
\end{equation}
which also corresponds to the minimum of the rate function \er{phi2}. The variance in turn reads, 
\begin{equation}
\frac{\langle K^2 \rangle - \langle K \rangle^2}{t} = - \left. \frac{d^2 \theta(s)}{ds^2} \right|_{s=0} = \frac{2}{9} \Omega
\; ,
\label{var2} 
\end{equation}

The whole of the SCGF and/or the rate function describes the shape of the overall distribution of $K$.  While \era{theta2}{phi2} resemble the form of the SCGF and rate function for a Poisson process, cf.\ \era{PKexLD}{Zsex}, there are small but crucial differences: in the quantum 2-level problem the total number of emissions is {\em sub-Poissonian}, as seen from the ratio of the variance to the mean, cf.\ \era{K2}{var2}, expressed as a Mandel-Q parameter,
\begin{equation}
Q = \frac{\langle K^2 \rangle - \langle K \rangle^2}{\langle K \rangle} - 1 = - \frac{2}{3} 
\; ,
\label{Q}
\end{equation}
where $Q=0$ for a Poisson process, and $Q$ positive (resp.\ negative) indicates super-Poissonian (resp.\ sub-Poissonian) statistics. Figure 10(b) shows that the rate function, and thus the distribution, is narrower than a Poisson rate function, and thus the statistics of $K$ displays smaller than a Poisson process. Quantum jumps are ``anti-bunched'' in time - i.e., events are anti-correlated. This is a quantum effect: immediately after a jump the system is in the $|0\rangle$ state, and coherent build up to the $|1\rangle$ state is needed before another jump can occur; this is manifested in the survival probability not being exponential, cf.\ \er{Pw2}.  In fact, the rate function \er{phi2} is that of a Conway-Maxwell-Poisson distribution \cite{Shmueli2005,Garrahan2010}.

While the behaviour of $\theta(s)$ around $s=0$ - or equivalently $\varphi(k)$ around its minimum - describes properties of {\em typical} dynamics, the behaviour of $\theta(s)$ for $s \neq 0$ encodes properties of atypical fluctuations of the dynamics.  The moments of the dynamical observable - such as the total number of jumps we are considering in this example - in the tilted ensemble of QJTs are
\begin{equation}
\langle K^n \rangle_s = \frac
{
\sum_{\omega_t} K^n(\omega_t) {\rm Prob(\omega_t)} e^{- s K(\omega_t)}
}
{
\sum_{\omega_t} {\rm Prob(\omega_t)} e^{- s K(\omega_t)}
}
\; , 
\label{Kn}
\end{equation}
where ${\rm Prob(\omega_t)}$ and the tilt is controlled by the counting field $s$.  At long times, the corresponding cumulants are obtained from the derivatives of $\theta(s)$. Figure 10(a) shows the tilted average $K$, 
\begin{equation}
k(s) = \lim_{t \to \infty} 
\frac{\langle K \rangle_s}{t} = - \theta'(s) = \frac{2}{3} e^{-s/3}
\; ,
\label{tps}
\end{equation}
for the 2-level example. We see that $k(s)$ goes from larger to smaller values as $s$ increases as expected, a $s < 0$ describes the more {\em active} than typical side of the dynamics, while $s > 0$, the more {\em inactive} side.  Figure 10(c) shows three sample trajectories taken from the $s=0$ (i.e., typical ensemble) for which $k(s=0)=2/3$ (center), and from the values of $s$ (i.e., tilted ensembles) for which $k(s) = 2$ (left) and $k(s) = 2/9$.  Higher derivatives of $\theta(s)$ describe properties of the fluctuations in the atypical QJT subensembles. An interesting observation is that, even if the average rate of emissions $k(s)$ changes with $s$, the properties of the fluctuations around the average, as measured by the $s$-dependent $Q$ parameter,
\begin{equation}
Q_s = \lim_{t \to \infty} 
\frac{\langle K^2 \rangle_s - \langle K \rangle_s^2}{\langle K \rangle_s} - 1 
= - \frac{\theta''(s)}{\theta'(s)} - 1 = - \frac{2}{3} 
\; ,
\label{Qs}
\end{equation}
are independent of $s$, describing a form of dynamical self-similarity in the 2-level system for this choice of parameters \cite{Garrahan2010}.

\begin{figure}[t]
	\includegraphics[width=\columnwidth]{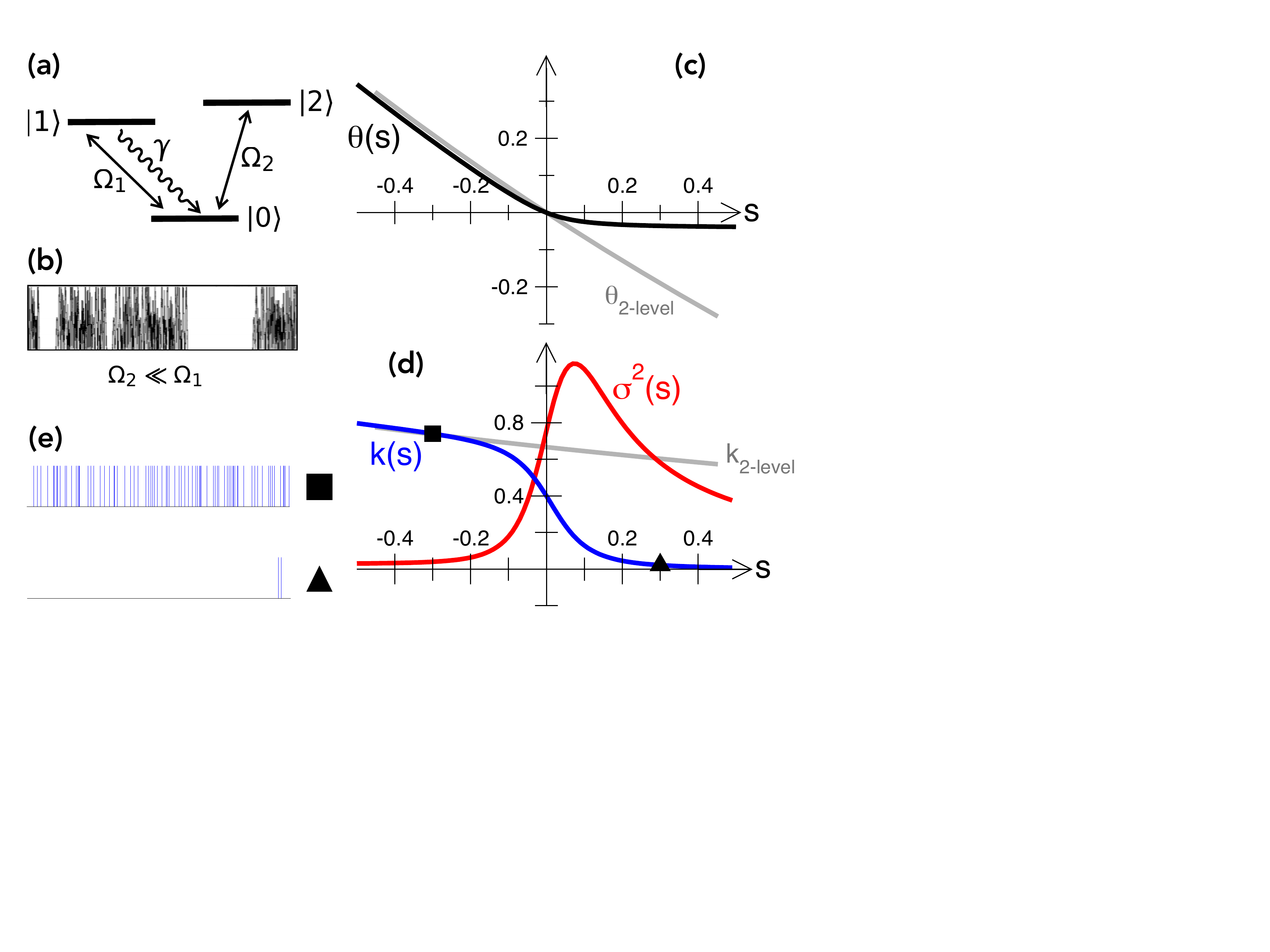}
\caption{
{\bf Large deviations in the 3-level system:}
(a) Level scheme of the 3-level system.
(b) Example of typical pattern of emissions. 
(c) SCGF as a function of $s$ (black), for $\Omega_2 = \Omega_1/10$ and $\gamma = 4 \Omega_1$, with the SCGF for the 2-level system (grey) for comparison.  From Ref.\ \cite{Garrahan2010}.
(d) Activity (blue) and variance (red) as a function of $s$.
(e) Examples of trajectories at various values of $s$.  
}
\label{fig11}
\end{figure}

\subsubsection{Example: 3-level system and intermittency}

As a second example we consider \cite{Garrahan2010} the 3-level problem of Fig.\ 11(a), by adding a third level $|2\rangle$ coherently coupled to level $|0\rangle$.  The Hamiltonian now reads,
\begin{equation}
H = \Omega_1 \left( | 1 \rangle \langle 0 | + | 0 \rangle \langle 1 | \right)
+ \Omega_2 \left( | 2 \rangle \langle 0 | + | 0 \rangle \langle 2 | \right) 
\; , 
\label{H3}
\end{equation}
while the single jump operator is given by
\begin{equation}
J = \sqrt{\gamma} | 1 \rangle \langle 0 |
\; .
\label{J3}
\end{equation}
For $\Omega_2 \ll \Omega_1$ this problem describes ``blinking'' dynamics, cf.\ Fig.\ 11(b), with a switching between periods of high emissions - corresponding to the system being in the subspace $| 0 \rangle|, | 1 \rangle$, and periods of no emissions when the system is ``shelved'' in state $| 2 \rangle$.  Such intermittent dynamics has a natural interpretation in terms of our ``thermodynamics of trajectories''. 

Figure 11(c) shows the SCGF for this problem (where we have chosen $\gamma = 4 \Omega_1$ and $\Omega_2 = \Omega_1/10$).  We see that for $s<0$ it tracks that of the 2-level system: atypically active dynamics is one where emissions have similar statistics to those of the 2-level system. In contrast, for $s>0$ the SCGF turns abruptly and becomes very flat. The corresponding average emission rate as s function of $s$ is shown in Fig.\ 11(d). For $s<0$ it has values similar to that of the 2-level system at the same $s$, and the trajectories look dense in jump events, cf.\ Fig.\ 11(e).  For $s>0$, $k(s)$ is close to zero and the trajectories are nearly empty of jump events, cf.\ Fig.\ 11(e). 
The order parameter $k(s)$ changes rapidly around $s=0$, and its variance displays a peak indicating enhanced fluctuations: this shows that there is a first-order crossover in the dynamics (a phase transition is not possible in such as finite sized system) between an active phase at $s<0$ and a largly inactive at $s>0$, with typical dynamics occurring at the crossover, and thus displaying strong fluctuations and intermittency.  This is the kind of ``thermodynamic'' interpretation of dynamical behaviour that the dynamical LD method permits.

\section*{Acknowledgements}
I am very grateful to my collaborators over the years in the joint work reviewed here, including R. Jack, I. Lesanovsky, V. Lecomte, F. van Wijland, S. Powell, M. Guta, B. Olmos, Y. Elmatad, L. Hedges, A. Keys, Z. Lan, E. Levi, K. Macieszczak, M. Merolle, E. Pitard, and M. van Horssen. Much of the work described in the early part of these notes was done together with my late friend and long-time collaborator David Chandler to whom this paper is dedicated. Financial support was provided by EPSRC Grant No. EP/M014266/1.


\bibliographystyle{elsarticle-num}
\bibliography{Brunico-1,Brunico-2,Brunico-3}

\end{document}